\documentclass[
reprint,
nofootinbib,
amsmath,amssymb,amsbsy,
prd,
onecolumn,
superscriptaddress,
longbibliography,
preprintnumbers]{revtex4-2}

\usepackage[T1]{fontenc}
\usepackage[shortlabels]{enumitem}
\usepackage{pifont}
\usepackage{amsmath}
\usepackage{xcolor}
\usepackage[mathscr]{euscript}
\newcommand{\mybar}[1]%
        {\kern 0.6pt\overline{\kern -0.6pt#1\kern -0.6pt}\kern 0.6pt}
\makeatletter
\newcommand{\subalign}[1]{%
  \vcenter{%
    \Let@ \restore@math@cr \default@tag
    \baselineskip\fontdimen10 \scriptfont\tw@
    \advance\baselineskip\fontdimen12 \scriptfont\tw@
    \lineskip\thr@@\fontdimen8 \scriptfont\thr@@
    \lineskiplimit\lineskip
    \ialign{\hfil$\m@th\scriptstyle##$&$\m@th\scriptstyle{}##$\hfil\crcr
      #1\crcr
    }%
  }%
}
\makeatother

 \usepackage{mathtools}
 \usepackage{lmodern}
 \usepackage{lmodern}
\usepackage{graphicx}
\usepackage{hyperref}
\usepackage{amsfonts}
\usepackage{tocvsec2}
\usepackage{slashed}
\usepackage{cancel}
\usepackage{comment}

\newcommand\beq{\begin{eqnarray}}
\newcommand\eeq{\end{eqnarray}}

\newcommand{\half}{\frac{1}{2}}

\newcommand\eqn[1]{\label{eq:#1}} 
\newcommand\eq[1]{eq.~(\ref{eq:#1})} 
\newcommand\eqs[2]{eqs.~(\ref{eq:#1},\ref{eq:#2})}

\newcommand{\bfr}{{\mathbf r}}

\newcommand{\GeV}{{\rm ~GeV }}

\newcommand{\MeV}{{\rm ~MeV }}
\newcommand{\CA}{{\cal A}}

\newcommand{\CC}{{\cal C}}

\newcommand{\CJ}{{\cal J}}

\newcommand{\CN}{{\cal N}}

\newcommand{\CL}{{\cal L}}

\def\msbar{\overline{MS}}
\def\amsbar{a\overline{MS}}

\def\half{\tfrac{1}{2}}

\def\U\Omega{U(1)_{\Omega}}

\hypersetup{
  colorlinks=true,   
  linkcolor=cyan,    
  citecolor=blue,    
  filecolor=magenta, 
  urlcolor=blue      
}
\usepackage[section]{placeins}

\begin{document}

\title{Dimer Effective Field Theory}

\author{Cullen Gantenberg}
\email{cgantenb@uw.edu}
\affiliation{Institute for Nuclear Theory, Box 351550, Seattle, Washington 98195-1550}
\author{David B. Kaplan}
\email{dbkaplan@uw.edu}
\affiliation{Institute for Nuclear Theory, Box 351550, Seattle, Washington 98195-1550}
\preprint{INT-PUB-26-010}

\begin{abstract}
 While chiral perturbation theory for mesons is characterized by a momentum expansion in $Q/\Lambda_\chi$ with  $\Lambda_\chi \approx 1\GeV$,  existing formulations of effective theory  for nucleon-nucleon scattering deviate  from data at $Q\approx 300 \MeV$ or lower.  We offer heuristic evidence  that  unsuspected nonanalytic structure exists in the complex momentum plane obstructing the effective field theory expansion in the spin-triplet channels, associated with the peak of the angular momentum barrier whose energy in low partial waves satisfies  $k=\sqrt{ME} \approx 300 \MeV$.  With this motivation, we construct a meromorphic function of $k^2$ we call the $C$-matrix, for which the radius of convergence of its Taylor expansion in $k^2$ is equivalent to that of the momentum expansion of the effective field theory. Thus the range of validity of the effective theory is directly related to the pole structure of the $C$-matrix. We uncover that pole structure and confirm that it is the source of the obstruction. The systematic  inclusion of dimer fields as propagating degrees of freedom in the effective theory to account for those poles   results in cut-off insensitive fits  at order $Q^0$ to most of the lower partial wave phase shifts up to the pion production threshold, using only the one pion exchange part of the long-range nucleon-nucleon interaction. Our theory should be applicable to the singular potentials regularly found in atomic physics as well.

\end{abstract}

\maketitle

\section{Introduction}

In developing successful theories of particle and nuclear physics, we are always limited by our ignorance about what is going on at short distance\footnote{For a recent compendium of papers on the status of our understanding of the nuclear interaction, see Ref.~\cite{tews2022nuclear}.}. 
If we include correctly all degrees of freedom  at  energies below some scale $\Lambda_\text{eft}$,  then by invoking Weinberg's ``folk theorem'' \footnote{{\it If one writes down the most general possible Lagrangian, including all terms consistent with assumed symmetry principles, and then calculates matrix elements with this Lagrangian to any given order of perturbation theory, the result will simply be the most general possible S-matrix consistent with perturbative unitarity, analyticity, cluster decomposition, and the assumed symmetry properties} \cite{weinberg_1979}.} we can represent the physics we have neglected by including all possible local interactions of these degrees of freedom consistent with symmetries and proportional to ``low energy constants'', which are then fit to data.  Each operator can be associated with a scaling dimension which renders most of them irrelevant to low energy physics. To use Wilson's terminology: when we compute  low energy observables, operators with coefficients that have large inverse mass scaling dimension will have to be accompanied by high powers of the light particle's momentum and hence will be small at low energy.  Therefore, at any order in a momentum expansion, only a finite number of operators will contribute, and with enough experimental data the theory can be predictive.  A momentum expansion of the scattering amplitude has a finite radius of convergence, however, set by the location of the lowest lying nonanalyticities in the complex momentum plane.  Such nonanalyticities arise from the propagation of physical particles that one has neglected in one's theory and will be characterized by the lowest particle threshold above $\Lambda_\text{eft}$.  To extend the range of validity of the local theory one has to include this new particle as an explicit degree of freedom in the effective theory. For example,  learning about the quantum numbers and symmetries of the  $W$, $Z$ and Higgs bosons  allowed the theoretical description of weak interactions to be extended to energies far beyond the limitations of Fermi's theory.

\begin{figure}[t]
    \centering
    \includegraphics[width=0.7\linewidth]{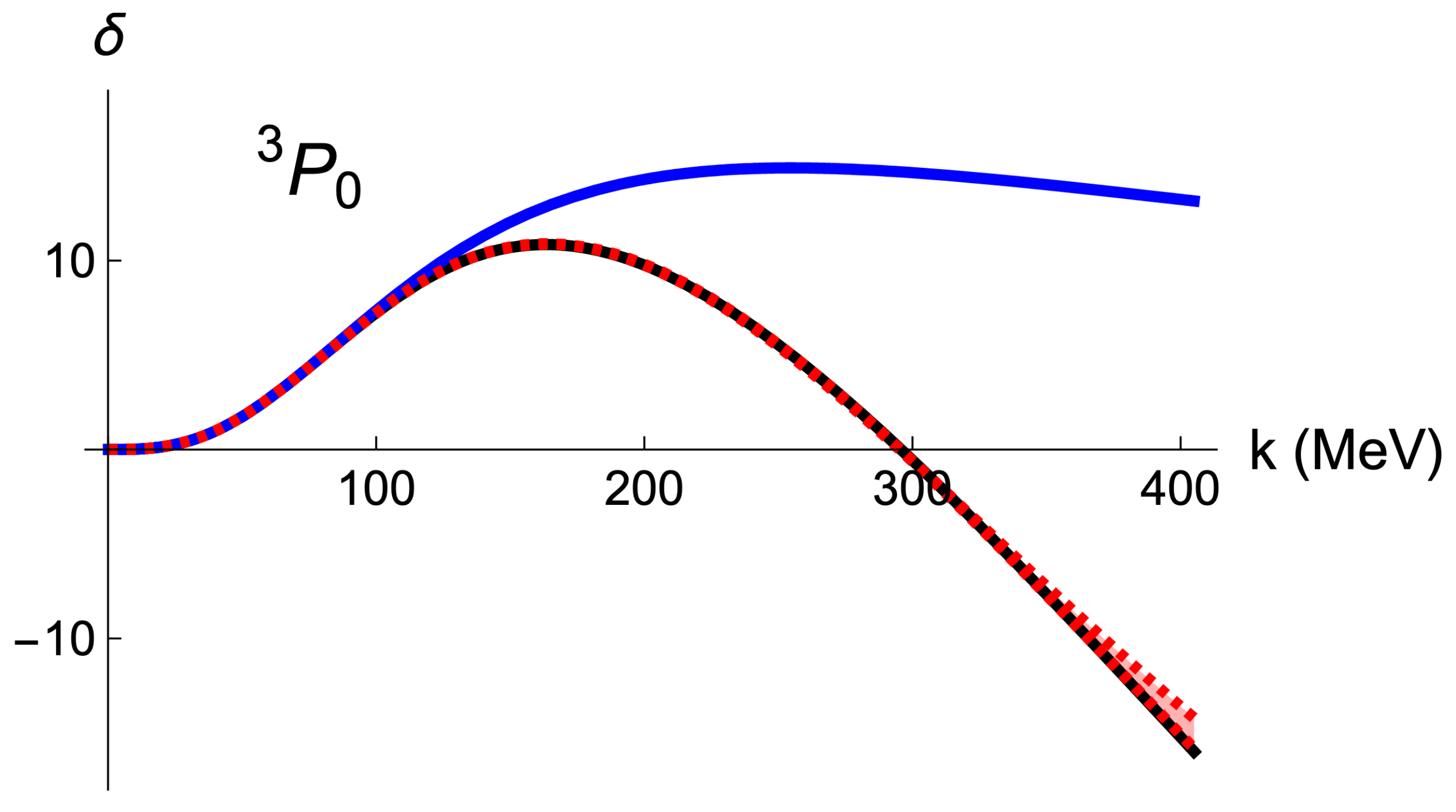}
    \caption{The ${}^3P_0$ nucleon-nucleon phase shift in degrees versus nucleon momentum in the center of mass frame ($p=405\MeV$ corresponds to $T_\text{lab} = 350\MeV$). The black curve is from the Nijmegen partial wave analysis \cite{Stoks:1993zz}, the blue curve indicates the effective range expansion fit to $O(p^4)$, and the dashed red curve is the result from the leading $O(Q^0)$ calculation in the dimer effective theory, which includes the one pion exchange potential.    The result depends weakly on the renormalization scale, the variation shown as this scale is varied through the range $\mu=100$ to $\mu=1500$ MeV. Details are provided in \S\ref{sec:NN}.}
    \label{fig:3P0phaseFinal.pdf}
\end{figure}

The topic of this paper is to explore whether the range of validity  can be similarly extended for the effective theory for nuclear interactions first proposed by Weinberg in 1990 \cite{Weinberg:1990rz,Weinberg:1991um,Ordonez:1992xp,ordonez1996two}.  This range is not as big as one might wish: while chiral perturbation theory for pseudoscalar mesons is controlled by an expansion in powers of $Q/\Lambda_\chi$, where $Q$ is a momentum or meson mass and $\Lambda_\chi\approx 1\GeV$ sets the UV scale, the effective theory for nucleon interactions seems to fail at a few hundred MeV in the spin-triplet channels for no obvious reason. Furthermore, this failure seems to entail an unexpected sensitivity to how one regulates the theory. Such sensitivity to UV physics undermines the rationale for an effective theory, where UV physics is supposed to be parametrized by low energy constants multiplying irrelevant operators, which are fit to data.   We will argue that these issues are due to nonanalyticities in the complex momentum plane of an object we call the $C$-matrix at a scale parametrically set by
$\Lambda_{NN} $, and proportional to $\ell^3$ for large angular momentum  $\ell$, where $\Lambda_{NN}$ is the scale introduced in Ref.~\cite{Kaplan:1996nv} and is defined by
\beq
 \Lambda_{NN}  = \frac{8\pi f_\pi^2}{g_A^2 M}\approxeq 300\MeV\ ,
\eeq
$f_\pi=132\MeV$ being the pion decay constant and $M$ the nucleon mass. 
Furthermore, we show how these nonanalyticities can be accounted for by including additional ``dimer'' fields  \cite{Kaplan:1996nv} to the effective theory,  restoring locality and extending the radius of convergence significantly.  As an example,   Fig.~\ref{fig:3P0phaseFinal.pdf}  shows the result from this dimer effective field theory for the phase shift for nucleon-nucleon scattering in the ${}^3P_0$ channel.  The black line is the phase shift in degrees as determined from the Nijmegen partial wave analysis of scattering data \cite{Stoks:1993zz}, plotted versus the nucleon momentum in the center of mass frame; the blue line is the result from using the first three terms in the effective range expansion. The dashed red curves are determined from the dimer effective theory at $O(Q^0)$ (which includes one pion exchange), where the band indicates the dependence on UV physics as parametrized by the regulator scale $\mu$ -- required to tame the singular tensor interaction -- as it is varied from $\mu=100\MeV$ to $\mu = 1500\MeV$. This result shows relative insensitivity to $\mu$ and a good fit up to $p=405\MeV$, corresponding to the lab kinetic energy $T_\text{lab}=350\MeV$.  For comparison, the Fermi momentum for nucleons in infinite matter at nuclear density is $k_F\approx 265\MeV$, so there is reason for optimism that this theoretical technology will allow for accurate computations for nuclei and nuclear matter, perhaps well beyond nuclear density.  The details for how we obtained  this result are found below in \S\ref{sec:NN}, along with the results for other partial waves (see Fig.~\ref{fig:AllPhasesV2}), but require some groundwork before we get there.

Weinberg's application of effective field theory to nuclear physics was very different from how effective field theory had been used up to that date.  By 1990, effective field theory had been applied previously to many processes, such as pion interactions,  weak interactions, neutrino masses, and proton decay, to name a few.   These are all feeble processes at low energy, and the explanation is that they arise from operators with dimension greater than four -- irrelevant operators.  Scattering or decay amplitudes  from these interactions will be proportional to powers of coupling constants with inverse mass dimension of size set by some relatively high mass scale (the QCD hadron resonance scale, the $W$ mass, the GUT scale for the examples given), and  powers of momenta   required in the numerator by dimensional analysis render these processes weak at low energy.  In contrast, nuclear interactions are called ``strong'' for a reason.  Nevertheless, at momenta below the pion mass, the only two-nucleon interaction one can write down in the ``pionless'' effective theory  is a  four-fermion vertex one would naively think was irrelevant.  However, Weinberg noted that such an interaction need not actually be irrelevant when the interacting particle is nonrelativistic, as the interaction will be enhanced by a power of the particle mass.   This is an effect familiar from quantum mechanics: no matter how weak an attractive potential is between two particles, if the particles are heavy enough they will form a bound state -- a nonperturbative phenomenon arising from a weak potential.  Although Weinberg considered the effective theory with pions, his reasoning also applies to the pionless theory where the leading two-nucleon interaction is expected to have strength $C_0 \approx g_A^2/f_\pi^2$ if dominated by one-pion exchange. A perturbative expansion of the scattering amplitude is found to take the form $C_0$ times powers of the ratio $Q/\Lambda_{NN}$.
 The larger the mass of the nucleon, the lower the scale $\Lambda_{NN}$ and the stronger the effect of the potential at fixed $Q$.  Weinberg's power counting scheme therefore treats $\Lambda_{NN} = O(Q)$, e.g. as an IR scale, so that the $C_0$ interaction must be treated nonperturbatively at leading order in the $Q$ expansion of the effective theory.  This summation is performed automatically when solving the Schr\"odinger equation with a $\delta$-function potential proportional to $C_0$, and so his prescription was to (i) compute the nucleon-nucleon potential to a given order in an expansion in $Q/\Lambda_\chi$ using chiral perturbation theory, and then (ii) solve the Schr\"odinger equation with this potential.  In this power counting scheme, the one-pion exchange potential is $O(Q^0)$, the two-pion exchange potential is $O(Q^2)$, while the leading contact interaction in a partial wave with angular momentum $\ell$ enters at $O(Q^{2\ell})$, as it requires $2\ell$ derivatives. A technical detail is that such potentials are generally highly singular and solving the Schr\"odinger equation requires invoking   a UV momentum cutoff $\Lambda_\text{UV}$  to render the calculation finite. (For calculations, the renormalization scale $\Lambda_\text{UV}$ need not be related to the scale $\Lambda_\text{eft}$ which determines the range of applicability of the effective theory.) One can renormalize up to the order in $Q$ to which one is working to absorb the cutoff dependence of the answer;  however, cutoff dependence of subleading contributions will remain, making it impossible to send $\Lambda_\text{UV}\to\infty$.  Nevertheless, the expectation in Weinberg's scheme is that as one works to successively higher order, the sensitivity to the cutoff will become successively weaker.  

     Problems with this expansion soon cropped up, such as those  pointed out by Nogga, Timmermans, and van Kolck \cite{Nogga:2005hy}.  These authors  showed that for $\ell>0$ at order $Q^0$, where the spin-triplet nucleon-nucleon potential is expected to be entirely due to one pion exchange, not only were the lower $\ell$  partial waves quite sensitive to the UV momentum cutoff $\Lambda_\text{UV}$, but that the cutoff dependence went through limit cycles causing the amplitude to diverge periodically at $p=0$ with increasing  $\Lambda_\text{UV}$.  However, they also showed that by introducing a four-fermion operator whose coefficient was tuned to the choice of cutoff, they could almost entirely eliminate the cutoff dependence. What is going on is clear:  the pion tensor force in the spin triplet channel scales as $1/r^3$ for small $r$, and an attractive $1/r^3$ potential sucks in one new bound state after another as the UV cutoff is raised -- being more singular than the kinetic term in the Schr\"odinger equation, which scales as $1/r^2$ -- causing poles in the scattering amplitude to progress up the imaginary momentum axis and causing the amplitude to blow up when the poles traverse the origin.  The contact interaction being promoted was serving to keep the poles away from the real axis, by preventing bound states via repulsion, or by sucking them deep into the potential via additional attraction. The troubling feature with this observation is that   such a contact term scales like $Q^{2\ell}$ and for $\ell>0$ has no business being used at the same order as one pion exchange, which enters at $O(Q^0)$.  The authors  discussed ``promoting'' the operator, redefining it to be $O(Q^0)$, which is equivalent to replacing $Q^{2\ell}/\Lambda_\chi^{2\ell}$ by $Q^{2\ell}/(\Lambda')^{2\ell}$ where $\Lambda'$ is some new IR scale taken to be $ O(Q)$, just like $\Lambda_{NN}$. However, if one does this, one must wonder which other operators should get promoted and one loses  the major virtue of effective field theory: that one can estimate systematic errors caused by truncating the calculation at some finite order in the $Q$ expansion.  Without a well defined power counting scheme, effective field theory simply becomes a model with unquantifiable uncertainties. 
     
    The periodic divergence of the threshold scattering amplitude is not the only problem encountered in that paper, even if the most  dramatic. In the ${}^3D_3$ channel, for example, even after promoting the counterterm by four orders in the expansion scheme and successfully eliminating the cutoff dependence,  the best fit obtained for the phase shift in Ref.~\cite{Nogga:2005hy} exhibits marked deviations from the data 
     at quite low momentum, below 100 MeV.  If this were due to the neglect of two-pion exchange interactions, one would have expected the deviations of the LO result from the data to scale like $(Q/\Lambda_\chi)^2$, or about 1\% at that scale,
instead of the significant deviation observed.   This corroborates that there is  an IR scale at play in the theory and that other operators are being promoted to higher relevance than expected in Weinberg's theory,  its effects  not being adequately accounted for by simply promoting the leading contact interaction.

Parallel and independent developments add to  this suspicion that some IR physics is missing from nuclear effective theories. Some years after Weinberg introduced his  theory, an alternative called the KSW expansion was proposed in Refs.~\cite{Kaplan:1998tg,Kaplan:1998we}, which involved the less ambitious expansion of the amplitude in powers of $Q/\Lambda_{NN}$, treating $\Lambda_{NN}$ as a UV scale and hoping it is sufficiently larger than $m_\pi$ for the expansion to be of use.  Such an expansion implies that pion exchange is to be accounted for perturbatively.  Two virtues of the KSW expansion is that  in the two-body sector  at least, amplitudes of interest can be computed analytically  and  can be fully renormalized, removing all dependence on a UV cutoff $\Lambda_\text{UV}$.    A momentum expansion of the amplitude is made compatible with nonperturbative physics (e.g. a very large scattering length in the ${}^1S_0$ partial wave, or the weakly bound deuteron in the ${}^3S_1$) by expanding around a nontrivial fixed point for the leading four-fermion interaction, endowing it with a large anomalous dimension so that it is no longer an irrelevant interaction. This idea was also discussed in \cite{van1999effective}. This fixed point corresponds to fermions with infinite $S$-wave scattering length (``unitary'' fermions).  All higher derivative operators receive large anomalous dimensions as well, and so are in effect ``promoted'' compared to their weak coupling analogs, a promotion precisely dictated by their renormalization group scaling.    The expected breakdown of the KSW expansion at $Q\approx \Lambda_{NN}$ occurs in the spin-singlet channel, but turns out not to be the case in the spin-triplet channel.  Fleming, Mehen, and Stewart showed  in Ref.~\cite{Fleming:1999ee} that NLO corrections indicated a precocious failure of the expansion in the lower partial waves, at $Q$ as low as $\approx \Lambda_{NN}/3$ in the ${}^3S_1$ channel.  They identified the problem as arising from ladder diagrams, and argued that the problem  would persist in the $m_\pi\to 0$  chiral limit.   

With the $1/r^3$ singularity in the spin-triplet channel once again being implicated as a source of trouble, Birse investigated solutions to the $1/r^3$ potential and showed that in any partial wave, as nucleon momentum increases, two real eigenvalues approach each other before merging and then separating again as imaginary complex conjugate pairs  \cite{Birse:2005um}.  He argued that this nonanalyticity would be an obstruction to the KSW expansion at a critical momentum $k_c$ that scaled proportionally to $\Lambda_{NN} \ell^3$, for large $\ell$.  He computed $k_c$ numerically for a number of the lower partial waves, however, and found that while parametrically proportional to $\Lambda_{NN}$, they had small numerical prefactors, which could explain the precocious breakdown of the KSW expansion at low momenta. Years later  in Ref.~\cite{Kaplan2020}  the ladder diagrams implicated in Ref.~\cite{Fleming:1999ee} for massless pion exchange in the spin-triplet channel were analytically computed up to seven loops, and indeed the breakdown of the KSW expansion they anticipated from a two-loop calculation was observed to persist, occurring at roughly the momentum value $k_c$ values determined by Birse.  

However, an unanswered question remained: what was the physical meaning of this  nonanalytic behavior uncovered by Birse?  
A clue may lie in a calculation performed in Ref.~\cite{Kaplan2020}.  In order to better understand the breakdown of the perturbative expansion for large $\ell$, the functional form of leading large $\ell$ behavior of the $n^{th}$ order perturbative contribution to the phase shift was extracted in the spin-triplet $L=J$ partial waves in the chiral limit, with the result for $n=1,\ldots 8$
\beq
  \frac{\delta^{(n)}}{\left(1+2(-1)^\ell\right)}\xrightarrow[]{\ell\to\infty}
 \left\{  \frac{\hat k}{\ell^2},
 \frac{3 \pi  \,\hat k^2 }{8 \ell^5}, 
 \frac{8 \,\hat k^3  }{3 \ell^8},
 \frac{315 \pi  \,\hat k^4  }{128 \ell^{11}},
 \frac{128  \,\hat k^5 }{5 \ell^{14}},
 \frac{15015 \pi   \,\hat k^6 }{512 \ell^{17}},
\frac{12288  \,\hat k^7 }{35 \ell^{20}},
\frac{14549535 \pi \,\hat k^8    }{32768 \ell^{23}},\ldots
 \right\}\ ,\cr &&
 \eqn{LeqJasympphase}\eeq
where $\hat k = k/\Lambda_{NN}$.  A functional form was then guessed which fit all eight of these terms,
 \beq
 \delta^{(n)} \sim \left(1+2(-1)^\ell\right)^n \frac{\hat k^n }{\ell^{3 n-1}}\frac{\sqrt{\pi }\, 2^{n-2}  \Gamma \left(\frac{3 n}{2}-\frac{1}{2}\right)}{\Gamma \left(\frac{n}{2}+1\right) \Gamma (n+1)}\ ,
  \eqn{Gam} \eeq
and this function was summed to all orders with the result
\beq
\delta&=& \sum_{n=1}^\infty \delta^{(n)}  
\cr &&
 =\frac{\hat k\left(2 (-1)^\ell+1\right)  }{\ell^2}\,  _3F_2\left(\frac{1}{3},\frac{2}{3},1;\frac{3}{2},\frac{3}{2};\frac{27 \hat k^2\left(2 (-1)^\ell
   +1\right)^2}{\ell^6}\right)-\frac{1}{2} \pi  \ell \left(\, _2F_1\left(-\frac{1}{6},\frac{1}{6};1;\frac{27 \hat k^2\left(2 (-1)^\ell
   +1\right)^2}{\ell^6}\right)-1\right)\ ,\cr &&
 \eeq
which exhibits a nonanalyticity in momentum when the argument of the hypergeometric functions exceeds one, namely for
   \beq
 \frac{27 \hat k^2\left(2 (-1)^\ell
   +1\right)^2}{\ell^6} \gtrsim 1\ ,\qquad \Longrightarrow\qquad   k \gtrsim  k_\star(\ell)\equiv \frac{\ell^3}{\sqrt{27}\, \left\vert 2 (-1)^\ell
   +1\right\vert}\,\Lambda_{NN}\ .
  \eqn{kstar} \eeq
\begin{figure}[t]
    \centering
    \includegraphics[width=0.4\linewidth]{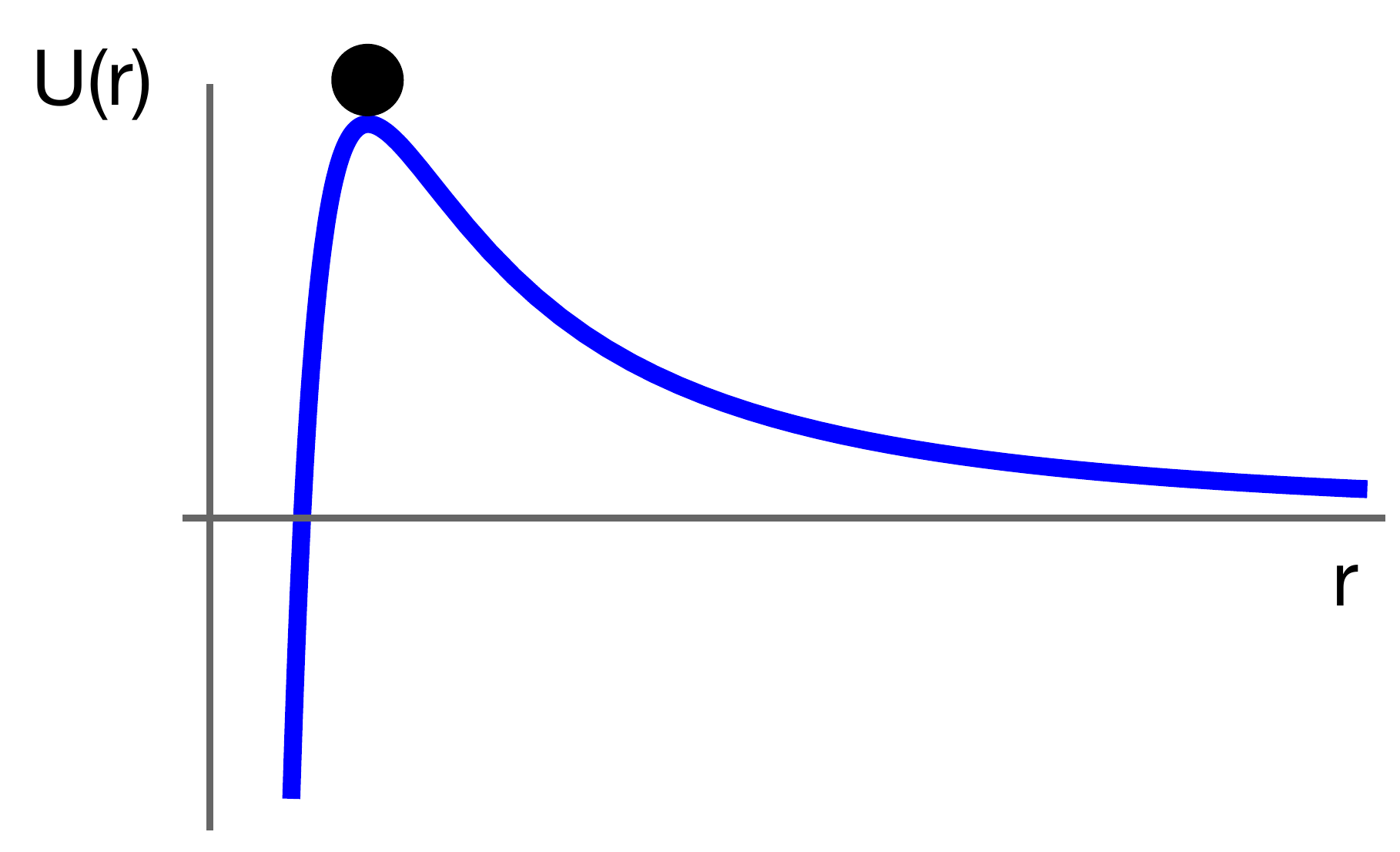}
    \caption{The stationary but unstable classical solution to a particle in an effective potential of form  $ U(r)=-g/r^3 + \ell^2/M r^2$  at large angular momentum $\ell$.}
    \label{fig:hump}
\end{figure}
The above result is consistent with the $\ell^3$ behavior anticipated by Birse but now with a precise coefficient (based on a guess) which can yield a clue as to its origin.  At large $\ell$ one might reasonably expect the system to behave classically, and so it is interesting to ask whether this nonanalyticity could be associated with a stationary solution to the classical  equations of motion. The reason that might be is that if there exists a stationary classical solution, the quantum mechanical particles might be likely to spend time in quantum states localized at short distance, causing a delay in the scattering and therefore a nonlocal looking scattering amplitude.  For the $L=J$ partial waves this corresponds to a particle in the effective potential $U(r) = -g/r^3 + \ell^2/M r^2$, where $g = 2\left[(-1)^\ell + 1\right]/(M\Lambda_{NN})$.  The unique stationary classical solution  for positive $g$ is the unstable one pictured in Fig.~\ref{fig:hump} where the particle sits at the top of the angular momentum barrier.  Remarkably, the energy of this solution is given by $E_{cl} = 4 \ell^6/(27M^3 g^2)$, corresponding  to a momentum $k_{cl}=\sqrt{M E_{cl}}$ which precisely matches the value of $k_\star$ given above in \eq{kstar}. 

This classical solution is very striking, and only possible in a potential more singular than $1/r^2$ at the origin.  It corresponds to a particle with positive energy trapped inside a potential that is negative everywhere, orbiting at a radius  $R=(\ell^2\Lambda_{NN})^{-1}$. A perturbation outward and the particle escapes; a perturbation inward and it plummets to the origin.  If a similar result holds for low $\ell$ then $R^{-1}\approx \Lambda_{NN}$ would correspond to a QCD scale that is heavier than the pion mass, yet lighter than the scale $\Lambda_\chi\approx 1\GeV$ associated with chiral symmetry breaking; it is an intermediate scale of great interest to nuclear physics and is close to the apparent failure point of the Weinberg effective field theory expansion\footnote{Treating $\Lambda_{NN}$ as an important IR scale  in the spin-triplet channel precludes a perturbative expansion about noninteracting nucleons;  Ref.~\cite{beane2002towards} proposed a novel perturbative expansion about the chiral limit instead, where at leading order the $1/r^3$ interaction from massless pion exchange is treated exactly.  However, we would expect that an expansion of the UV physics around the chiral one pion exchange potential would still see a limited radius of convergence due to the resonance structure inherent in the $1/r^3$ potential, which is insensitive to the pion mass.}. 

This may seem like a lot of weight to put on a flimsy chain of reasoning, but it does motivate considering the possibility that in quantum mechanical scattering from a singular potential there may be nonanalyticities in the complex momentum plane which exist at a radius $\approx k_\star$ from the origin.  In this paper we pursue this idea, we find that indeed such nonanalytic structures do exist, and we develop the machinery for including their effects in a local effective theory by adding additional, propagating ``dimer'' fields\footnote{These dimer fields have baryon number 2, and are not related to the $\Delta$ resonance; it also seems unlikely that the dimer is related to a $\Delta\Delta$ intermediate state, such as was treated in Ref.~\cite{savage1997delta}, given that such a state should be significantly heavier.}. The result is in uniformly better agreement with two-body scattering data up to significantly higher momentum than typically seen in prior effective field theory treatments \footnote{For earlier attempts to use dimers to improve results from the Weinberg expansion, see \cite{habashi2022nucleon,Sanchez2018two}.}.

To pursue this idea we begin by addressing the question of how to extend the range of validity for a nonrelativistic and nonperturbative effective field theory,  starting with simple examples  and building up to the realistic case of interest in several stages.  In the next section we discuss constructing a local effective theory for particles with purely short distance interactions, successively accounting for not only virtual and weakly bound states, but also resonances. This is like the pionless effective theory for nucleon interactions, and consists entirely of contact interactions with any number of derivatives. The first step is to identify exactly what quantity has a momentum expansion that is in one-to-one correspondence with the derivative expansion of the effective Lagrangian.  This is what we call the $C$-matrix, which for the simplest examples is closely related to what is called the $K$-matrix, the Cayley transform of the $S$-matrix.  This is a meromorphic function in the complex momentum plane  whose Taylor expansion coefficients are directly related to the coupling constants in the effective theory.  Thus the radius of  convergence of the momentum expansion of the effective theory is governed by the location of the closest pole to the origin in the $C$-matrix.  We show that one can systematically incorporate dimer fields in the effective theory (fields with fermion number two exchanged in the $s$-channel \cite{Kaplan:1996nv})  to effectively remove these poles and extend the convergence of the derivative expansion to arbitrarily high momentum scales.

We next turn to the more interesting case where there are both short- and long-range interactions in the full theory, but begin by restricting our attention to long-range potentials that are less singular than $r^{-2}$ at the origin.    We again show how to construct the $C$-matrix in this case, which is now no longer simply related to the $K$-matrix but combines both the scattering data and certain nonperturbative quantities one can compute from the form of the long distance potential.   Once again this $C$-matrix is meromorphic and the range of validity of the momentum expansion of the effective theory can in principle be extended to arbitrarily large momentum, given adequate scattering data.

Finally we consider the case relevant for nucleon scattering, with an emphasis  on the spin-triplet channels, where  both short- and long-range interactions  coexist, but the separation between the two is ambiguous due to the latter being more singular than $1/r^2$ near the origin.  Here we show how to obtain similar results described in the previous section, but with UV-scale dependent couplings for both the dimer and nucleon fields, conspiring in the $C$-matrix to reproduce the data, which knows nothing of that scale. The challenge lies in how to find poles in the $C$-matrix in the complex momentum plane when we are constrained to data on the real axis.  We do not simply introduce a lot of new parameters to fit, but address this problem by exploiting the fact that we expect to see interesting features at distance scales where the interaction is dominated by pion physics, where we have an analytical handle on the interaction.  This allows us to identify and locate the poles and residues of interest to a fair degree accuracy using a combination of analytical and numerical methods before performing a fit to data, which we find  adjusts their values by $\lesssim 30\%$. 

We conclude with a brief mention of the spin-singlet channels, three-body forces and trimers, and then end with a discussion section outlining the things that need improved understanding.

\begin{figure}[t]
    \centering
    \includegraphics[width=0.5\linewidth]{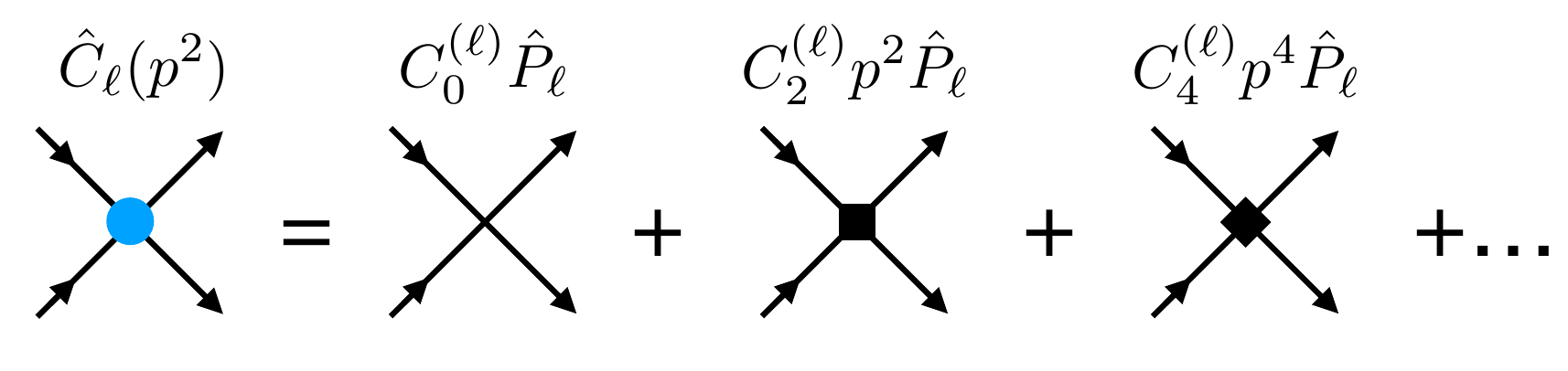}
    \caption{{\it Tree level contact interaction in the effective theory for scattering in the angular momentum $\ell$ partial wave, where $\hat P_\ell$ is a $2\ell$-derivative operator that projects both incoming and outgoing pairs into the angular momentum $\ell$ state}}
    \label{fig:EFTvertex}
\end{figure}

\section{The dimer effective field theory  for purely short distance interactions}

\subsection{Defining the $C$-matrix}

 We first consider the ``pionless'' effective field theory, appropriate when  interactions are all at a length scale $\Lambda^{-1}$ or shorter.  The effective theory  will then consist entirely of contact interactions in a derivative expansion.  Such interactions are automatically separable with the appropriate renormalization scheme, and the two-body scattering amplitude can be  computed nonperturbatively. We begin by reviewing the discussion of such a system as given in Ref.~\cite{Kaplan:1996nv,Kaplan:1998tg,Kaplan:1998we}.
 We take the tree level vertex  in the center of mass frame for the $\ell^\text{th}$ partial wave to all orders in the momentum expansion (as shown in Fig.~\ref{fig:EFTvertex}) to be 
\beq
\text{tree level vertex } = -i
p^{2\ell}\sum_{n=0}^\infty C^{(\ell)}_{2n}  p^{2n}\equiv  
-i 
p^{2\ell}C_\ell(p^2)\ , 
\eqn{tree}
\eeq
where the coefficients $C^{(\ell)}_{2n}$ are various linear combinations of the
operator coefficients in the Lagrangian  contributing to $\ell$-wave scattering with mass dimension $-2(\ell+1+n)$, while $M$ and $p$ are the mass and the magnitude of the momentum for each of the incoming fermions in the center of momentum frame\footnote{Here for simplicity we ignore  spin and other quantum numbers such as  isospin -- they are easily incorporated into the formalism.}. Contact operators can be represented in the Lagrangian  in terms of Galilean invariant combinations of spatial gradients, eliminating time derivatives by making use of the free equations of motion\footnote{The equations of motion can be used even for internal vertices of a Feynman diagram since their effect is to shrink to a point the fermion propagator they act on, essentially creating a new contact interaction.  Since all possible contact interactions are being included in the effective theory anyway, this will not affect the generality of the theory.}.   The factor of $p^{2\ell}$ arises from an  operator $\hat P_\ell$ at the four-fermion vertex which projects incoming and outgoing fermion pairs into the angular momentum $\ell$ state, and whose normalization is fixed by \eq{tree}.   

\begin{figure}[t]
    \centering
    \includegraphics[width=0.8\linewidth]{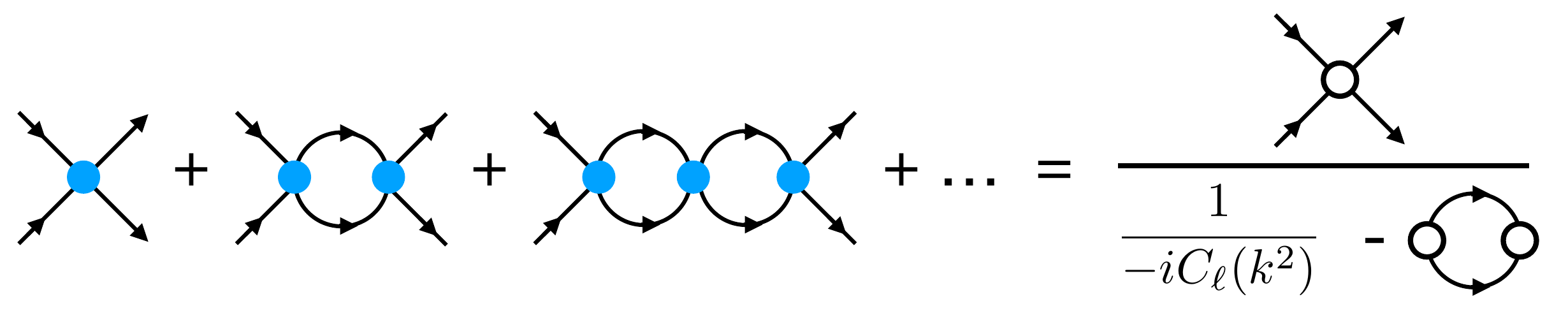}
    \caption{{\it The complete sum of Feynman diagrams for 2-body scattering arising from local operators, incorporating the entire two-body interaction from the ``pionless'' effective theory to all orders in momentum. The blue circles  represent   the 4-fermion contact interaction $-i\hat C_\ell(p^2)$ as shown in Fig.~\ref{fig:EFTvertex} containing $\hat P_\ell$ projection operator, while $C_\ell(p^2)$ is the vertex polynomial defined in \eq{tree}.    The empty circles correspond to just the projection operator without any $-iC$ coupling. }}
    \label{fig:bubbles}
\end{figure}

Since there are no long distance interactions in this effective theory by assumption, and the contact interactions comprise a separable potential, the exact Feynman scattering amplitude in the nonrelativistic limit may be computed as the sum of bubble diagrams shown in Fig.~\ref{fig:bubbles}, and  is in general given by  \cite{Kaplan:1998tg,Kaplan:1998we}
\beq
i\CA^{(\ell)}_\text{eft} 
& = & \frac{ k^{2\ell} 
}{
\frac{1}{-iC_\ell(k^2)} +  i\left(F+i\frac{M }{4\pi} k^{2\ell+1}\right)  }  = i \frac{4\pi}{M}\frac{k^{2\ell}}{\left[-\left(\frac{4\pi}{M}\right)\left(\frac{1}{C_\ell(k^2)}+F\right) -ik^{2\ell+1})\right]}
\ , 
\eqn{examp}
\eeq 
where $k\equiv\sqrt{ME}$ is the magnitude of the on-shell momentum.  In the above expression the  $ik^{2\ell+1}$ term is the scheme-invariant nonanalytic contribution from the two-fermion cut,  $C_\ell(k^2)$ is the renormalized vertex,  and  $F$ is a power series in $k^2$ of dimension $\text{(mass)}^{2\ell+2}$ with finite coefficients, which defines the renormalization scheme used for computing the bubble diagrams in  Fig.~\ref{fig:bubbles}. Different choices of $F$ corresponding to  finite shifts of the $C_{2n}$ coefficients;  in the $\msbar$ scheme, for example,   one defines $F=0$, and so $F$ represents the difference between $\msbar$ and the chosen subtraction scheme.   A momentum cutoff scheme, for example, would have $ F$ proportional to the cutoff raised to the appropriate power.

In principle we can measure the scattering phase shifts and determine from them the $C$ coefficients of the contact interactions in the effective theory. To that end we invert the  above expression to obtain
\beq
C_\ell(k^2) = -\left[ i\frac{M}{4\pi}   k^{2\ell+1}  + \frac{k^{2\ell}}{\CA^{(\ell)}_\text{eft}}+ F\right]^{-1}\ .
\eqn{Ceft}
\eeq
Since the goal of the effective theory is to have $\CA^{(\ell)}_\text{eft}$ closely approximate the exact amplitude $\CA^{(\ell)}$ at low momenta, we can replace  $\CA^{(\ell)}_\text{eft}\to \CA^{(\ell)}$ and use data to determine $\CA^{(\ell)}$.   Therefore it follows that the coefficients of the contact interactions which we want to determine are given by the  momentum expansion in powers of $k^2$ of what we call the $C$-matrix, defined as
\beq
 {\mathbf C_\ell}\equiv 
-\left[ i\frac{M}{4\pi}   k^{2\ell+1}  + \frac{k^{2\ell}}{\CA^{(\ell)}}+ F\right]^{-1}= - \left[
\frac{M k^{2\ell+1}}{4\pi}\cot\delta_\ell+F
\right]^{-1}=
 \left[
\hat K_\ell^{-1}-F
\right]^{-1}
\ ,
\eqn{Cmat1}\eeq
where we used the relation between the $\CA_\ell$ and the phase shift $\delta_\ell$
\beq
\CA^{(\ell)} = \frac{4\pi}{M} \frac{k^{2\ell}}{k^{2\ell+1}\cot \delta_\ell - i k^{2\ell+1}}\ ,
\eqn{Adelt}\eeq
and defined the modified $K$-matrix as\footnote{Our definition of $\hat K_\ell$ differs   by a factor of $-  4\pi/(M k^{2\ell+1})$ from the conventional definition of the $K$-matrix as the Cayley transform of the $S$-matrix.}, 
\beq
\hat K_\ell \equiv
-\left({\rm Re}\left[\frac{k^{2\ell}}{\CA_\ell} \right]\right)^{-1}
= - \frac{4\pi}{M k^{2\ell+1}}\tan\delta_\ell\ .
\eqn{Khat}
\eeq

We call the quantity ${\mathbf C}$  in \eq{Cmat1} the ``$C$-matrix''.  It is a function of $k^2$ which depends on the underlying physics through the phase shift, which can in part be determined experimentally. However, unlike the $S$-matrix or $K$-matrix, it is renormalization scheme dependent through the choice of  $F$, which is finite.  Two schemes we employ are $\msbar$ $(F=0)$ and what we call the $\amsbar$ scheme,
\beq
F = \lim_{k\to 0} \hat K^{-1}_\ell\ .\qquad (\amsbar)
\eqn{Fdef1}
\eeq

The utility of the $C$-matrix lies in its   direct connection with the low energy constants of the effective theory even when the interactions are nonperturbative, unlike more physical quantities such as the quantity $p^{2\ell+1}\cot\delta$, which is subject to the effective range expansion.  Since  the renormalized $C^{(\ell)}_{2n}$ operator coefficients equal the Taylor series coefficients when the $C$-matrix is expanded in powers of $k^2$, the radius of convergence for any physical prediction from the effective theory will be limited by the distance from the origin in the complex $k$ plane to the nearest nonanalyticity of the $C$-matrix.

The effective range expansion exploits the fact that when there are only short range interactions, the quantity $k^{2\ell+1}\cot\delta_\ell$ is an even function of $k$ and is analytic for $|k|\lesssim \Lambda$ with a Taylor expansion in powers of $k^2$  of the form
\beq
k^{2\ell+1}\cot\delta_\ell = -\frac{1}{a} + \half r_0 k^2 + \textstyle{\frac{1}{4}}r_1 k^4+\ldots\ ,
\eqn{ere}
\eeq
 where $a$ is the scattering length and $r_0$ is the effective range for $\ell=0$, or their analogs for higher $\ell$. Each of these parameters is naturally of order $\Lambda$ to the power dictated by their dimension.  However, the interaction can be fine-tuned to have a weakly bound or an almost bound state, in which case the parameter $a$ can  be parametrically larger than its generic value $\Lambda^{-(2\ell+1)}$.  Being essentially an expansion in powers of $ k^2/\Lambda^2$, 
 the effective range expansion would appear to be  closely related to the derivative expansion in the effective field theory, \eq{tree}.  As we have seen, though, the $C$-matrix is not proportional to the quantity $k^{2\ell+1}\cot\delta_\ell $  but instead depends on its inverse, the $\hat K$-matrix, a meromorphic function that can have  poles arbitrarily close to the origin, depending on the choice of subtraction scheme $F$.  This does not appear to be consistent with   the definition of our vertex defined in \eq{tree} plus the assumption that the effective theory is valid up to $k\approx \Lambda$.  For example, if  $k^{2\ell+1}\cot\delta_\ell $ is well approximated by the quadratic polynomial $(-1/a + r_0 k^2)$ and we work in the $\msbar$ scheme with $F(p^2)=0$, then the $C$-matrix has  poles at $k=\pm(a r_0)^{-1/2}$, which will be at $|k|\ll\Lambda$ when $a\gg \Lambda^{-1}$ -- namely, when there is a low-lying bound or virtual state.  This can complicate the task of representing low energy physics via an effective field theory, an issue addressed in Ref.~\cite{Kaplan:1996xu}.  

 Such a low-lying pole is not a big problem, however, because of the freedom to choose $F$.  In Weinberg's scheme, where one employs a momentum cutoff, one typically has $F\approx \Lambda$, while in the KSW scheme one takes $F$ to equal the momentum scale one is interested in, and in either case the pole in the $C$-matrix is pushed out far enough from the origin to not be a problem.
We will see, however, that there can be additional poles in the $C$-matrix that are harder to avoid, in particular ones at a scale of $\Lambda_{NN}\approx 300\MeV$  in the spin-triplet partial waves.  Furthermore, we would like to have a systematic way to at least in principle extend the range of validity of the effective theory to as high a momentum as we wish.  This is a feature well understood in perturbative, relativistic effective field theories but which has had no counterpart in the nonrelativistic, nonperturbative effective theories we are interested in here.   
In an effective theory such as Fermi's theory of weak interactions, the radius of convergence for the momentum expansion is set by the distance  from the origin in the complex momentum plane to the nearest nonanalyticity of the scattering amplitude which is not properly accounted for  by the effective theory.  These nonanalyticities are due to propagating particles that have not been included -- such as the $W$, $Z$, and  Higgs bosons, in the example of Fermi's theory -- which give rise to poles and cuts in lepton and hadron scattering amplitudes.  To extend the range of validity of the theory, a UV extension was constructed to properly account for these heavier degrees of freedom -- namely the $SU(2)\times U(1)$ gauge theory for weak interactions.  This theory then contains irrelevant operators to explain neutrino masses, as well as potentially proton decay, electric dipole moments, etc, motivating the search for ``BSM'' models which could extend the UV completion to even higher scales.   In a nonrelativistic, nonperturbative theory it was a long-standing problem  how perform similar UV extensions, the problem being how to  incorporate in the same theory  without double counting both interacting constituents as well as fundamental degrees of freedom for their bound states   \cite{weinberg1963elementary,weinberg1963quasiparticles}. The theory of dimers was devised to solve exactly this problem in Ref.~\cite{Kaplan:1996nv}. Here we extend its application beyond bound and virtual states to include resonances.  As we will demonstrate, this application of dimers allows one to systematically extend the range of the effective theory as high as one wants by including ever more dimers.

\subsection{The dimer expansion}
\label{sec:dimexp}

  In perturbative, relativistic theories  new particles show up as poles or cuts in the scattering amplitude as the momenta is increased above the particle production threshold.  One immediately sees that it cannot work like that in nonperturbative, nonrelativistic  theories, however. For low-lying bound and virtual states in two-body $S$-wave scattering with a large scattering length the scattering amplitude at low momentum looks approximately like
  \beq
 \CA\approx \frac{4\pi}{M} \frac{1}{\left(-1/a - i k\right)}
 \eeq
 -- which has a simple pole at $k = i/a$ -- but  $k =\sqrt{ME}$ is nonanalytic in the energy (arising from a 2-particle cut due to the constituents of the composite state), and such a pole in the amplitude cannot be accounted for by the tree level  exchange of a particle in the $s$-channel in the way that the $Z$ boson can account for a pole in    the $s$-channel $e^+e^-$ scattering amplitude.  
 In Ref.~\cite{Kaplan:1996nv} it was shown that the correct way  to account for such a nonanalyticity is to incorporate a light, propagating degree of freedom (the ``dimer'') whose the tree-level propagator approximates the $C$-matrix, rather than the amplitude, which is a meromorphic function of $k^2$.  The idea is to include fields in the effective theory with fermion number two, and incorporate their effects into the local four-fermion vertex as shown in Fig.~\ref{fig:bubbles}.  Once the dimer has accounted for the low-lying poles in the $C$-matrix (with the correct residues), the remainder will be analytic out to the first remaining pole, and that radius will determine the radius of convergence for the derivative expansion.  Although the dimer theory of Ref.~\cite{Kaplan:1996nv} has been applied to diverse  physical systems \cite{hammer2020nuclear,epelbaum2009modern,bedaque1999renormalization,bertulani2002effective,hammer2004universal,schwenk2005resonant,ando2005effective,nishida2012impossibility,briceno2013three,brown2014field,hale2014effective,kamiya2016structure,hammer2017effective,epelbaum2021effective,ando2023s,esposito2025short}, a consistent dimer expansion has not been systematically formulated to date, nor has it been extended beyond low-lying bound and virtual states to include resonances.  
 
\begin{figure}
    \centering
    \includegraphics[width=0.75\linewidth]{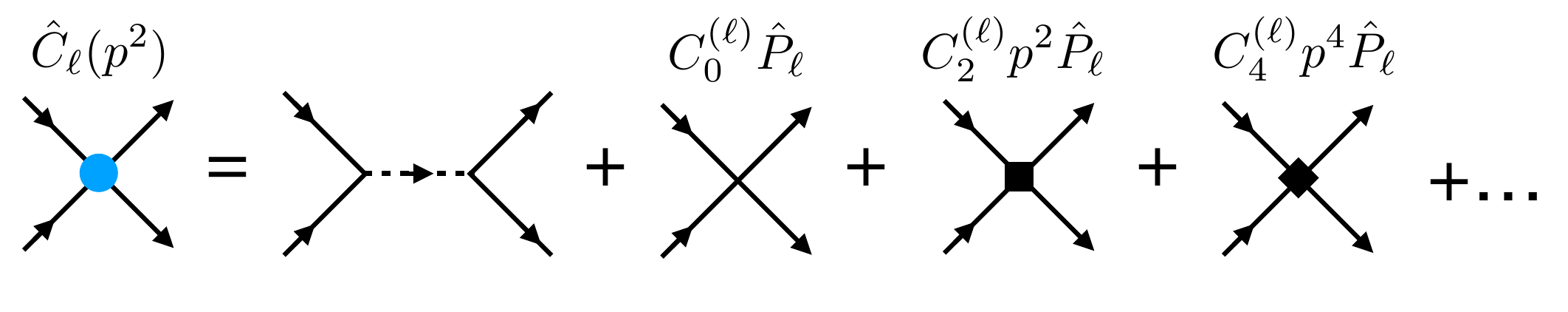}
    \caption{\it In the dimer theory, the vertex appearing in Fig.~\ref{fig:bubbles} is augmented to include the exchange of one or more $s$-channel dimers.}
    \label{fig:DEFTvertex}
\end{figure}

  To see how this works, we will designate a pole as occurring at $k=k_\star$ with residue $Z$.  Because the $C$-matrix is both real and   even in $k$, these poles will generically appear as quartets $\{\pm k_\star, \pm  k^*_\star\}$.  Furthermore, the residues at the poles obey a similar set of reflection properties.   
Thus, for example, a quartet of poles appears in the $C$-matrix in the combination
\beq
\left[\frac{Z}{k-k_\star} - \frac{Z}{k+k_\star}+ \frac{  Z^*}{k-  k^*_\star} - \frac{  Z^*}{k+  k^*_\star}\right] = \frac{  y^2}{E-E_\star} + \frac{    {y^*}^2}{E-E^*_\star}\ ,\qquad E_\star = \frac{k_\star^2}{M}\ ,\quad y^2 = \frac{2 Z k_\star}{M}\ .
\eqn{dimquart}
\eeq
When combined like this, the poles look like a local interaction arising from the propagators of a pair of dimers in the $s$-channel, where the two dimer fields can be included in the Lagrangian as
\beq
\CL_\text{dimer} = \Phi_1^\dagger\left(i\partial_t - \frac{\nabla^2}{4M}- E_\star\right) \Phi_1  + \frac{y}{2}\left( \Phi_1 \psi^\dagger \psi^\dagger+ h.c.\right) + \Phi.c.\ , 
\eqn{Ldimer}
\eeq
where ``$\Phi.c.$'' refers ``$\Phi$-conjugation'',  an operation entailing simultaneous hermitian conjugation and $\Phi_1\leftrightarrow\Phi_2$ exchange. Note that while the Lagrangian is not real (since $y$ is in general complex), this $\Phi$-hermiticity ensures that the $C$-matrix is real and hence that the theory is unitary.  It is key that dimers do not appear as physical states in the Hilbert space but only as intermediary particles that allows us to represent the Lagrangian in terms of local interactions.  This is in turn will allow us to gauge the theory and study the interactions with photons, for example.

For the case where $k_\star$ is either real or imaginary, the poles must come in $\{\pm k_\star\}$ pairs and the pole structure of the $C$-matrix  takes the form
\beq
\left[\frac{Z}{k-k_\star} - \frac{Z}{k+k_\star}\right] = \frac{y^2}{E-E_\star}\ ,
\eeq
with the same definitions of $E_\star$ and $y^2$, except now $Z$ is real if $k_\star$ is real, and imaginary if $k_\star$ is imaginary.
The single dimer Lagrangian takes the same form as the two-dimer Lagrangian except without the ``$+\Phi.c.$'' since it is self $\Phi$-conjugate in this case with the definition $\Phi\to \pm \Phi^\dagger$, which choice of sign depending on the reality of $y$.

A technical  issue with including dimers is that, while poles will generically be at a distance $O(\Lambda)$ from the origin, if the scattering length is large there can be low-lying poles which introduce a low-lying mass scale to keep track of. We can use the scheme dependence of the $C$-matrix to avoid this by choosing the finite $F$ subtraction to move the low-lying pole to the origin, achieved by implementing \eq{Fdef1}.  This is what we call the ``scattering length subtraction scheme'', or $\amsbar$ for short.

At this point it is simplest to work out a toy example with solvable UV physics to illustrate how the dimer expansion works in a solvable model.  To that end we consider $S$-wave scattering from a $\delta$-shell potential.  

\subsubsection{Example: A primary dimer for $S$-wave scattering by an attractive $\delta$-shell potential}
\label{sec:dshell1}

\begin{figure}[t]
    \centering
    \includegraphics[width=0.45\linewidth]{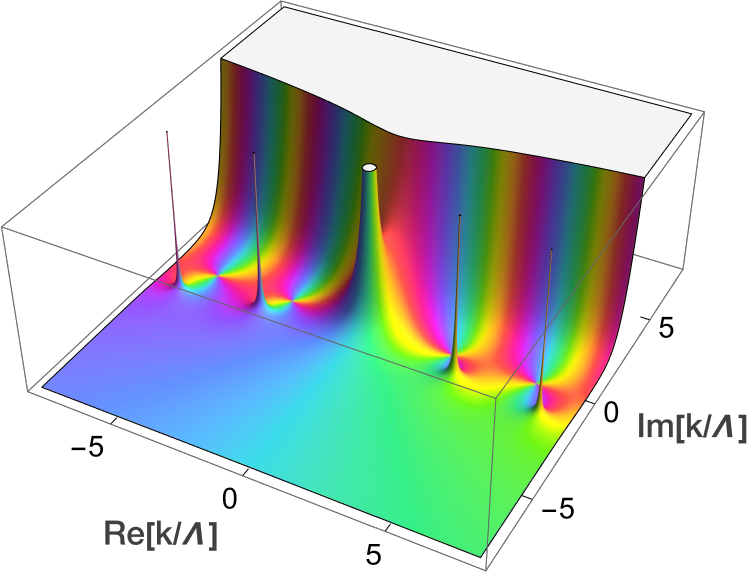}
    \caption{{\it The $\ell=0$ scattering amplitude for the $\delta$-shell potential with $g=0.9$ and radius $1/\Lambda$ in the complex $k$ plane, exhibiting both a pole on the negative imaginary axis (the virtual bound state) as well as a series of poles  below the real axis (long-lived resonances corresponding to the particle rattling around inside the shell).} }
    \label{fig:ComplexAmplitude_dshell}
\end{figure}

The $\delta$-shell potential  with a  radius $1/\Lambda$  given by
\beq
V(r) =  -\frac{g\Lambda}{M}\delta\left(r-\frac{1}{\Lambda}\right)
\eeq
is a convenient toy model for the underlying UV physics that the effective theory is trying to describe because it is exactly solvable, short range, and yet not so singular that it requires renormalization.   The solution yields the scattering amplitude
\beq
\CA_\ell = \frac{4 \pi  g j^2_\ell(\xi )}{\Lambda  M \left(-i g \xi  j^2_\ell(\xi )+g \xi  j_\ell(\xi ) y_\ell(\xi )+1\right)}\ ,\qquad \xi \equiv \frac{k}{\Lambda} = \frac{\sqrt{ME}}{\Lambda}\ ,
\eeq
where $j_\ell$, $y_\ell$ are spherical Bessel functions (see, for example, Ref.~\cite{gottfried2018quantum}). 
Fig.~\ref{fig:ComplexAmplitude_dshell} shows a plot  of the scattering amplitude for $\ell=0$ and $g = 0.9$ in the complex $\xi$ plane. Obvious features are the pole on the negative imaginary axis at $\xi =-0.1\, i $, as well as a sequence of poles in the lower half plane at $\xi = (\pm 3.7 - 1.1\, i),\, (\pm 6.9 -1.4 \,i),\ldots$

From the amplitude one can compute
\beq
 k^{2\ell+1}\cot\delta_\ell = \frac{\Lambda ^{2 \ell+1} \xi ^{2  \ell} (g \,\xi  j_ \ell(\xi ) \,y_ \ell(\xi )+1)}{g j^2_ \ell(\xi )}&\equiv &
 -\frac{1}{a} + \sum_{n=1}^\infty \frac{r_n}{n} k^{2n}\ ,
\eqn{ERE}
\eeq
where the first few terms in the effective range expansion are given by
\beq
-\frac{1}{a} &=&\frac{2^{2 \ell+1} (-g+2 \ell+1) \Lambda ^{2 \ell+1} \Gamma \left(\ell+\frac{1}{2}\right) \Gamma
   \left(\ell+\frac{3}{2}\right)}{\pi  g}\ ,\cr &&\cr
   r_0 &=& \frac{2^{2 \ell+3} (-g+2 \ell-1) \Lambda ^{2 \ell-1} \Gamma \left(\ell+\frac{3}{2}\right)^2}{\pi  g
   (4 \ell (\ell+1)-3)}\ ,\cr &&\cr 
   r_1 &=& -\frac{4^{\ell+2} (\ell+3) (g-2 \ell+3) \Lambda ^{2 \ell-3} \Gamma \left(\ell+\frac{3}{2}\right)^2}{\pi 
   g (2 \ell-3) (2 \ell+3)^2 (2 \ell+5)}\ ,\cr &&\cr 
 r_2&=& -\frac{2^{2 \ell+3} (\ell (2 \ell+15)+30) (g-2 \ell+5) \Lambda ^{2 \ell-5} \Gamma
   \left(\ell+\frac{3}{2}\right)^2}{\pi  g (2 \ell-5) (2 \ell+3)^3 (2 \ell+5) (2 \ell+7)}\ .
  \eqn{EREc} \eeq
The modified $K$-matrix is given by
\beq
\hat K_\ell = -\frac{4\pi}{M\Lambda k^{2\ell}}\frac{ j_\ell(\xi )^2}{    \xi  j_\ell(\xi ) y_\ell(\xi )+\frac{1}{g} }\ .
\eqn{KUV}
\eeq
The subtraction constant $F$ is then defined as in \eq{Fdef1} as
\beq
F = \lim_{p\to 0}\hat K^{-1} =M  \Lambda ^{2 \ell+1} \frac{2^{2 \ell-1} (g-2 \ell-1)\Gamma \left(\ell+\frac{1}{2}\right) \Gamma \left(\ell+\frac{3}{2}\right)}{\pi ^2 g} \ ,\qquad (\amsbar\text{ scheme})
\eeq
which yields the $C$-matrix using \eq{Cmat1}, 
\beq
{\mathbf C}_\ell  = \left[ \frac{1}{\hat K_\ell} - F\right]^{-1} = \left[
-\frac{M\Lambda}{4\pi}\left( k^{2\ell} 
\frac{    \xi  j_\ell(\xi ) y_\ell(\xi )+\frac{1}{g} }{ j_\ell(\xi )^2}
+   \Lambda ^{2 \ell} \frac{2^{2 \ell+1} (g-2 \ell-1)\Gamma \left(\ell+\frac{1}{2}\right) \Gamma \left(\ell+\frac{3}{2}\right)}{\pi  g}\right)\right]^{-1} \ .
\eqn{CmatUV}\eeq
\begin{figure}[t]
    \centering
    \includegraphics[width=.9\linewidth]{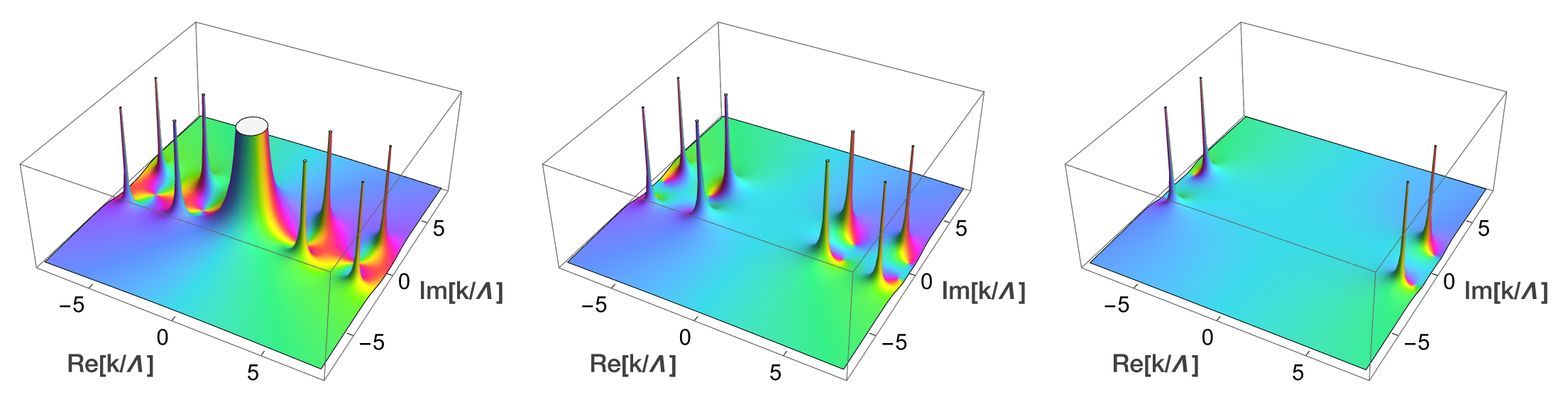}
    \caption{{\it In the left figure we plot the $\ell=0$ $C$-matrix for the $\delta$-shell potential with $g=0.9$ and radius $1/\Lambda$ in the complex $k/\Lambda$ plane.  On the left is the $C$-matrix  computed in the $\amsbar$ scheme, setting  $F= 1/a$ in \eq{examp}, exhibiting a $1/\xi^2$ pole at the origin.   This pole at $\xi=0$ corresponds indirectly to the virtual state pole in the amplitude (Fig.~\ref{fig:ComplexAmplitude_dshell}), while the outer quartets of poles in the $C$-matrix correspond to the pairs of resonance poles in the amplitude.  In the center frame we plot the $C$-matrix with the $1/k^2$ pole subtracted, corresponding to the primary dimer field contribution, exhibiting a radius of convergence for the momentum expansion out to the first resonance quartet.   On the right we plot the $C$-matrix with both the primary dimer and the first resonance dimer contributions subtracted, and we see that the radius of convergence for the momentum expansion  has been extended out to the second resonance quartet.}}
    \label{fig:Cmatplot}
\end{figure}

 In the left panel of Fig.~\ref{fig:Cmatplot} we plot the $C$-matrix in the complex momentum plane for $\ell=0$ and $g=0.9$, exhibiting its reflection symmetry about both axes. The pole at the origin is associated with the virtual state and arises from the $C_{-2}$ term; the quartets of poles away from the origin are associated with the pairs of resonance states, but their locations do not correspond to poles in the amplitude. 
We can  extract the $C_{2n}$ coefficients, starting at $n=-1$, by a Laurent expansion of the $C$-matrix in powers of momentum.  The first few coefficients in the Laurent expansion about $\xi=0$ given by
\beq
 C_{-2}^{(\ell)} = -\frac{4\pi}{M}\frac{2}{r_0} \ , \qquad C^{(\ell)}_0 = \frac{4\pi}{M}\frac{r_1}{r_0^2}\ ,\qquad C^{(\ell)}_2 =  -\frac{8\pi}{M} \left(\frac{r_1^2}{4 r_0^3}-\frac{r_2}{3 r_0^2}\right) \ .
 \eeq 
where we have expressed the results in terms of the effective range parameters from \eq{EREc}, rather than as functions of $g$ and $\Lambda$. Note that none of these coefficients look fine tuned:  the critical coupling in the $\ell=0$ partial wave where the scattering length is $g_c=1$, so that the level of fine tuning in this model is parameterized by $1/(g-g_c) = 10$ for our choice of $g=0.9$, but of all the effective range expansion coefficients, only the scattering length depends on this number. By working in the $\amsbar$ scheme we have eliminated all dependence of the $C$-matrix on the scattering length, and therefore none of the dimer or contact operator parameters will be fine-tuned.

\begin{figure}[t]
     \centering
    \includegraphics[width=0.6\linewidth]{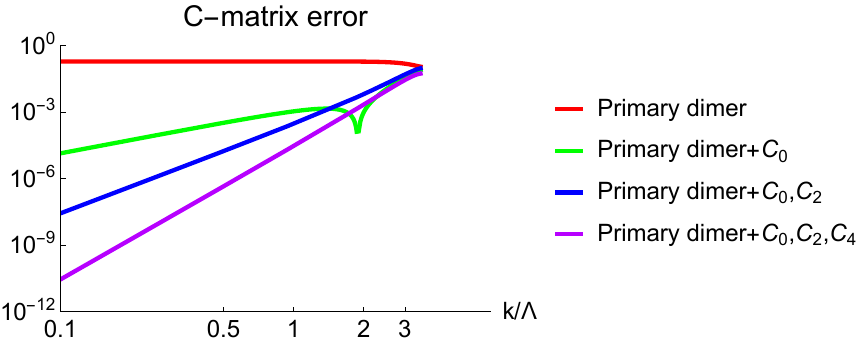}
     \caption{\it The Lepage plot for the error incurred in truncation of the $C$-matrix when employing the primary dimer plus contact interactions, showing the improvement that results from including higher derivative interactions.  The plot makes it evident that the radius of convergence for this effective theory is at $k\approx 3\Lambda$. }
     \label{fig:PrimaryDimerLepage}
\end{figure}

In the dimer effective theory,  the  $C^{(\ell)}_{-2}$ term is represented by the tree-level  exchange of one or more dimers (\eq{Ldimer}) while the $C_{2n}^{(\ell)}$ terms for $n\ge 0$ correspond to  4-fermion contact interactions with $2n$ derivatives in addition to the $2\ell$ derivatives require by the projection operators to the $\ell$ partial wave, as pictured in Fig.~\ref{fig:DEFTvertex}.  Those contact interactions are therefore directly the Taylor expansion coefficients for the $C$-matrix with the central pole removed, shown in the central panel in  Fig.~\ref{fig:Cmatplot}.  One can see that the radius of convergence of this expansion will be  set by  the innermost pole quartet, associated with the first resonance.  A Lepage plot \cite{Lepage:1997cs} for the error incurred at truncating the dimer expansion of the $C$-matrix at the first few orders is displayed in Fig.~\ref{fig:PrimaryDimerLepage}, where we have used a single dimer field (the ``primary dimer'') to represent the central pole at the origin of the $C$-matrix in the $\amsbar$ scheme, 
demonstrating a radius of convergence of $\approx 3\Lambda$.

\subsubsection{Example: Extending the radius of convergence by including a resonance dimer pair}

\begin{figure}[t]
    \centering
    \includegraphics[width=0.65\linewidth]{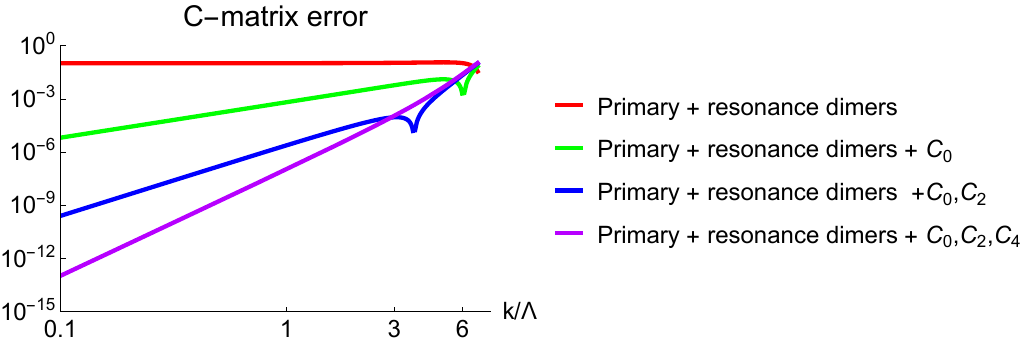}
     \caption{\it The Lepage plot for the error incurred in truncation of the $C$-matrix when employing both the primary dimer and the first resonance dimers, plus contact interactions. Here one sees that by including the resonance dimers we have extended the radius of convergence for the effective theory to $k\approx 6 \Lambda$ while also improving the prediction for the phase shift at lower momentum for the same number of contact interactions.}
     \label{fig:SecondaryDimerLepage}
\end{figure}

Now it should be obvious how to proceed to expand the radius of convergence -- one merely needs to include more dimers to account for successive pole quartets in the $C$-matrix.  In particular, we need to numerically solve for the pole position $k_\star$ and residue $Z$ to be plugged into \eqs{dimquart}{Ldimer}.
For the present model of $S$-wave scattering off the $\delta$-shell potential with $g=0.9$ 
we find a pole location $\xi_\star  = (3.76+1.43\,i)$ and residue $Z=(0.115 -0.046\,i) $.  When we include such a dimer pair for the innermost pole quartet, the Feynman graphs for this theory still take the form in Fig.~\ref{fig:DEFTvertex}, but now the first graph involves a sum over all three dimers -- the primary dimer associated with the $C$-matrix's pole at the origin, and the dimer pair associated with the innermost pole quartet. 
The $C_{2n}^{(\ell)}$ coefficients for $n\ge 0$ now correspond to the Taylor expansion coefficients for the $C$-matrix with the central pole and the innermost pole quartet removed, shown in the right panel of Fig.~\ref{fig:DEFTvertex}. The radius of convergence for the expansion is now set by the second pole quartet in the $C$-matrix, associated with the second resonance of the amplitude. The Lepage plot  in Fig.~\ref{fig:SecondaryDimerLepage} shows that by including the first resonance dimer pair in addition to the primary dimer, the radius of convergence for the derivative expansion has been extended from $k\approx 3\Lambda$  to $k\approx 6\Lambda$. It is clear that this process of extending the range of validity of the effective field theory by adding additional dimers can be continued, while increasing the accuracy of the effective theory for a given radius of convergence can be accomplished by including successively higher dimension contact interactions. Furthermore, it is also evident that since the $C$-matrix is being reproduced to ever better accuracy without the introduction of any new and spurious nonanalyticities, the double counting problem that concerned Weinberg in Refs.\cite{weinberg1963elementary,weinberg1963quasiparticles}-- where one has both fundamental and composite states where there should only be one -- is not an issue in the dimer effective field theory approach -- no new poles have been introduced in the amplitude.
 
\begin{figure}[t]
    \centering
    \includegraphics[width=1\linewidth]{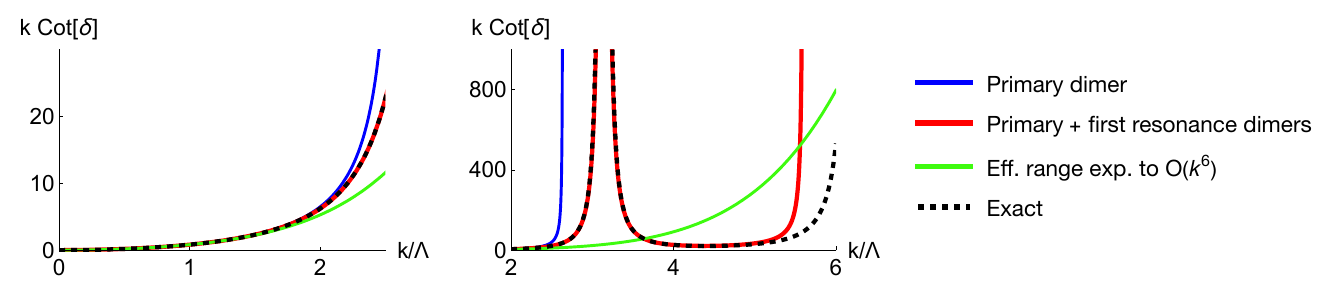}
    \caption{{\it Plots of  $k\cot\delta$ for two different ranges of $k/\Lambda$  for $\ell=0$ scattering via the $\delta$-shell potential with coupling $g=0.9$ and radius $1/\Lambda=1$.  The exact result is given by the black dashed line.  The effective range expansion (green) computed to $O(k^6)$ exhibits deviations from the exact result at $k/\Lambda\approxeq 1.5$.  The theory with one dimer and contact interactions to $O(k^2)$ does somewhat better (blue line) than the effective range expansion with the same number of free parameters. When an additional dimer pair is added to reproduce the lowest lying resonance pole quartet in the $C$-matrix, one sees a dramatic increase in the radius of convergence of the momentum expansion (red).} }
    \label{fig:pcd2dimerplt}
\end{figure}

To close this section we also provide the prediction of the dimer effective field theory for a quantity directly accessible to experiment, the phase shift.  In Fig.~\ref{fig:pcd2dimerplt} we plot $k\cot\delta$ for the $\delta$-shell potential in the dimer theory with (i) just the primary dimer (blue) and (ii) with the primary dimer and the first resonance dimer pair (red).  In both cases we expand the 4-fermion contact interactions out to $C_2 k^2$.    For comparison we also plot the effective range expansion to order $k^6$ (solid green) and the exact analytic result (black, dashed)\footnote{The dimer expansion with just the primary dimer  involves fitting the scattering length and effective range to fix $\amsbar$ subtraction point and dimer coupling respectively, and then there are two additional fit parameters $C_{0,2}$ for a total of four.  The effective range expansion to $O(k^6)$ similarly involves four parameters: $a$, $r_{0,1,2}$.}. This plot shows that the theory with a primary dimer is somewhat better than the effective range expansion, although roughly equivalent, while inclusion of the resonance dimers allows the validity of effective theory to be extended far beyond the effective range expansion, correctly accounting for the lowest scattering resonance.

The conclusion from this example is that the effective theory for a short range potential can have its radius of convergence for the momentum expansion extended without limit by adding dimers to provide the poles found in the $C$-matrix out to the desired radius.  This is a bit counter-intuitive -- for the present example it means that for a $\delta$-shell potential with spatial size  $1/\Lambda$, we can in principal construct an effective theory which describes scattering accurately up to momentum a million times $\Lambda$, all in terms of dimers. (Ignoring relativity, of course!) We will use this fact when we tackle the problem of greatest interest, nucleon-nucleon scattering in the triplet channel.


\section{Dimer effective theory in the presence of long range interactions}

\subsection{Generalizing the  $C$-matrix}
\label{sec:KCIR}

\begin{figure}[t]
    \centering
    \includegraphics[width=0.7\linewidth]{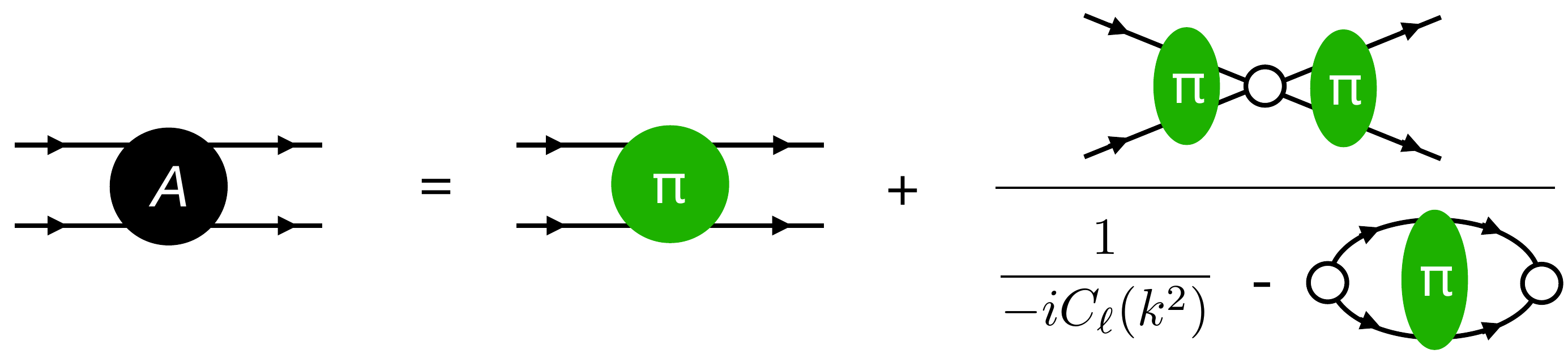}
    \caption{\it The analog of the expansion in Fig.~\ref{fig:bubbles} for the ``pionful'' effective theory  when long distance interactions are included (e.g., pion exchange). The   empty circles corresponds to a $\hat P_\ell$ projection operator, while $C_\ell(p^2)$ is defined as in \eq{tree}.  The full amplitude is the sum of 1-particle irreducible diagrams, the blobs marked ``$\pi$'' involving all possible insertions  of the long-range interactions  (containing no contact or dimer-mediated interactions) which leave the graph connected. }
    \label{fig:pibubbles}
\end{figure}

\begin{figure}[b]
    \centering
    \includegraphics[width=0.3\linewidth]{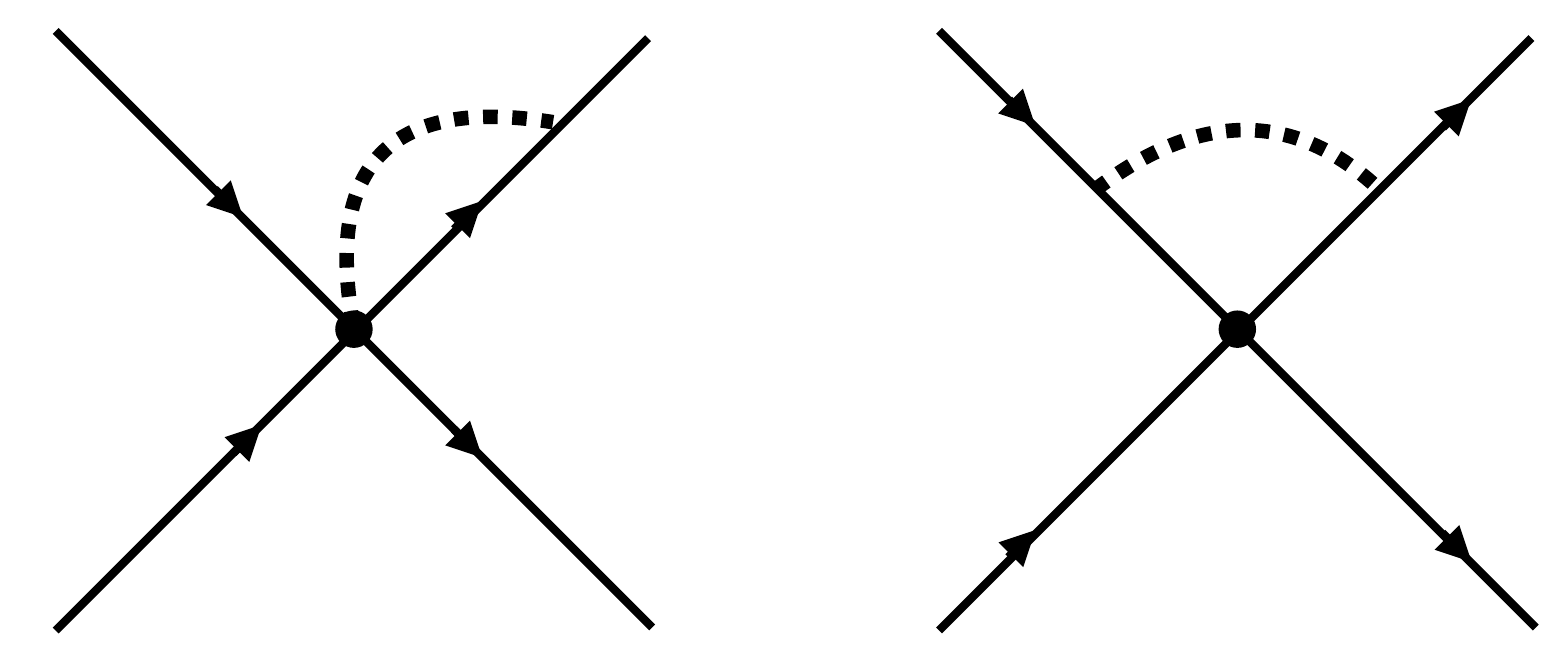}
    \caption{\it Radiation pion contributions which are neglected in the sum of diagrams of Fig.~\ref{fig:pibubbles}.  See Ref.~\cite{Fleming:1999ee} for a discussion. }
    \label{fig:RadPi}
\end{figure}

For a theory of nucleon interactions we have not only short distance (UV) physics, but long distance (IR) physics due to the exchange of pions.  The fact that the contact interactions in the effective theory are separable, even  though long distance interactions are not, allows us to represent the 2-body scattering amplitude nonperturbatively by the infinite sum of diagrams from a local theory as shown in Fig.~\ref{fig:pibubbles}, generalizing the diagrams of Fig.~\ref{fig:bubbles} that we encountered in the theory without long distance interactions\footnote{We do not consider here graphs involving ``radiation pions'',  such as those in Fig.~\ref{fig:RadPi}, which involve a new scale $Q_r = \sqrt{M m_\pi}$ -- even though eventually they should be accounted for, especially as one approaches the pion production threshold.  See Appendix~C of Ref.~\cite{Fleming:1999ee} for a discussion.}.

We need to turn Fig.~\ref{fig:pibubbles} into a nonperturbative definition for the amplitude that can be used for computations. We start by considering the IR physics by itself:  a particle with reduced mass $M/2$, energy $E= k^2/M$, and angular momentum $\ell$,  scattering from the  potential $V_\text{IR}(\bfr)$ which is assumed to be central and to be regulated so that it is less singular than $r^{-2}$ at the origin.  The wave function $\chi_\ell(k,r)$ satisfies the Schr\"odinger equation
\beq
  \left[\frac{\partial^2}{\partial r^2}+\frac{2}{r}\frac{\partial}{\partial r}+k^2-\frac{\ell(\ell+1)}{r^2}-MV_\text{IR}(r)\right]\chi_\ell=0\ .
\eqn{ SchIRL}
\eeq
There will be two independent solutions to this equation.  One possible  pair  are the Jost solutions $f_\ell^\pm(k,r)$, with $f_\ell^-(k,r)\equiv \left[f_\ell^+(k,r)\right]^*$, which are defined by their asymptotic behavior at large $r$:
\beq
    f_\ell^\pm(k,r)\xrightarrow[]{r\to\infty}\frac{e^{\pm ikr}}{kr}\ .
\eqn{Jost}
\eeq
Alternatively,  we can define independent solutions $\CJ_\ell(k r)$ and $\CN_\ell(k r)$ by their  behavior at small $r$:
\beq
    \CJ_\ell(k,r)\xrightarrow[]{r\to0}\frac{(kr)^{\ell}}{(2\ell+1)!!}\ ,\qquad
    \CN_\ell(k,r)\xrightarrow[]{r\to0}-\frac{(2\ell-1)!!}{(kr)^{\ell+1}}\ ,
\eqn{JNnorm}
\eeq
where the normalizations have been chosen to match the $r\to 0$ behavior of the regular and irregular spherical Bessel functions $j_\ell$ and $n_\ell$ respectively.  
The $\CJ$ and $\CN$ solutions can then be related to the Jost solutions by
\beq
    \CJ_\ell(k,r)=\left(yf_\ell^+(k,r)e^{-i\pi \ell/2}+c.c.\right)\ ,\qquad
    \CN_\ell(k,r)=\left(zf_\ell^+(k,r)e^{-i\pi \ell/2}+c.c.\right)\ ,
\eqn{JNJost}\eeq
where the  complex numbers $y$ and $z$ must be determined by solving the Schr\"odinger equation.  From \eqs{Jost}{JNnorm} one can compute the Wronskians 
\beq
    W(f_\ell^+,f_\ell^-)=-\frac{2i}{kr^2}\ ,\qquad    W(\CJ,\CN)=\frac{1}{kr^2}\ ,
\eqn{Wronski}
\eeq
which imply the relation between the $y$ and $z$ coefficients
\beq
    y^*z-yz^*=-\frac{i}{2}\ .
\eqn{Wyz}
\eeq
Without UV interactions, the only acceptable scattering solution up to normalization is  
\beq
\Psi_\ell^\text{IR} =\CJ_\ell(k,r)\ ,
\eeq
by virtue of it being regular at the origin. By using \eq{JNJost} we can then express the corresponding scattering matrix in terms of $y$ as
\beq
S^\text{IR}_\ell =e^{2i\delta_\text{IR}} = -\frac{y\ }{y^*}\ ,
\eeq
and from this the $K$-matrix
\beq
\hat K_\ell^\text{IR} =- \left({\rm Re}\left[\frac{k^{2\ell}}{\CA^\text{IR}_\ell}\right]\right)^{-1} 
=\frac{4\pi}{Mk^{2\ell+1}} \frac{\,\text{Re}[y]}{\,\text{Im}[y]} \ ,
\eqn{KIR}\eeq
where $\CA_\ell  = 2\pi/(M k) (S _\ell-1)$

In the effective theory where we have both contact interactions and the $V_\text{IR}$ potential, the wave function away from the origin is given instead by
\beq
\Psi_\ell=a \CJ_\ell + b \CN_\ell\ ,
\eqn{Psi}
\eeq
with the $a$ and $b$ coefficients determined by the contact terms.  The reason why we can't have such a wavefunction without the contact terms 
is because the Wronskian $W(\Psi^*,\Psi) \propto 1/r^2$ implies that the probability flux through a small sphere about the origin of radius $R$ does not vanish as $R\to 0$, meaning that the IR Hamiltonian by itself is not self-adjoint if we allow such singular wave functions.  However, when there are UV interactions, the role of contact interactions in the effective theory is to allow solutions such as these, with real, energy dependent $a,b$ coefficients.  One can either think of the contact interactions as altering the boundary conditions at the origin, or modifying the Hamiltonian to make it self-adjoint.  

The scattering parameters corresponding to the above wave function $\Psi_\ell$ are  determined by the asymptotic behavior of the $\CJ$ and $\CN$ solutions given in \eq{JNJost}:
\beq
S_\ell = -\frac{a y + b z}{a y^* + b z^*} &=&
-\frac{y\ }{y^*} -\left(\frac{1}{y^*}\right)^2\frac{z y^* - y z^*}{a/b +z^*/y^*}
=S_\ell^{\text{IR}}+\frac{1}{(2y^*)^2}\frac{2i}{a/b + z^*/y^*}\ ,
\eqn{SUV}
\eeq
where we used \eq{Wyz}.  The scattering amplitude is then given by
\beq
i\CA_\ell = \frac{2\pi}{Mk}(S-1) = i\CA_\ell^{\rm IR} +\frac{4\pi }{M k} \frac{1}{(2y^*)^2}\frac{i}{a/b + z^*/y^*}\ .
\eeq

Next we need to understand how the ratio $a/b$ is related to the contact interactions. It is important to recognize that while the $\CJ_\ell$ and $\CN_\ell$ functions -- and the $y,z$ coefficients that govern their asymptotic behavior -- are derived entirely from $V_\text{IR}$, a nonzero ratio $b/a$ is due entirely to the $C$ coefficients parameterizing the UV physics, since the contact interactions only care about the singularity structure at the origin.  In fact, if we kept the UV physics but turned off the IR physics, the wave function \eq{Psi} would turn into what we will call $\Psi_\ell^\text{UV}$ which for the same contact interactions is given by the same wave function but with $\CJ$ and $\CN$ replaced by ordinary spherical Bessel functions, the free solutions with the identical behavior near the origin:
\beq
\Psi_\ell^\text{UV} = a j_\ell + b n_\ell\ .
\eeq

For the Bessel functions we have $y\to -i/2$ and $z\to -1/2$, and so
\beq
S_\ell^\text{UV} = \frac{a-ib}{a+ib}\ ,\qquad  \CA^\text{UV}_\ell = -\frac{4\pi}{Mk} \frac{1}{a/b +i}\ ,\qquad \hat{K}^\text{UV}_\ell = \frac{4\pi}{M k^{2\ell+1}}\frac{b}{a}\ .
\eeq

However, we have already computed this same amplitude in \eq{examp} with the result
\beq
i\CA_\ell^\text{UV} = i\frac{4\pi}{M}\frac{k^{2\ell}}{-\frac{4\pi}{M}\left(\frac{1}{C_\ell(k^2)} + F\right) - i k^{2\ell+1}}\ .
\eeq
On equating these two expressions we obtain
\beq
\frac{a}{b} = \frac{4\pi}{M k^{2\ell+1}}\left(\frac{1}{C_\ell(E)}+F\right)
\eqn{abc}
\eeq
and  can rewrite the scattering amplitude in the effective theory as
\beq
i\CA_\ell = i\CA_\ell^\text{IR} - \frac{ k^{2\ell}/(2 y^*)^2}{i\left(\frac{1}{ C_\ell(E)} +F\right)+ i \frac{M}{4\pi} k^{2\ell+1} \frac{z^*}{y^*}}\ .
\eqn{master}
\eeq

It is possible to now see a one-to-one correspondence between the terms in the above expression and the graphs on the right side of Fig.~\ref{fig:pibubbles}.  One major difference is that while the individual graphs are divergent and therefore the $C_\ell$ vertex in the graphs is the bare one,  our derivation of the amplitude in \eq{master} started with the physical scattering solution, and therefore  the terms appearing in it are finite renormalized quantities, with the function $F$ being the only reminder that there is scheme dependence to the solution.  Of course, if $V_\text{IR}$ has to be regulated to make it less singular than $1/r^2$ at the origin, then the $y$ and $z$ coefficients will depend on this regulator, and it remains to be seen whether the theory can be renormalized as that regulator is removed.  However, all discussion of infinities arising from  the singularity of the contact interactions have been avoided.

 From \eq{master} it is convenient to derive the modified $K$-matrix
\beq
\hat K_\ell = -\left({\rm Re}\left[k^{2\ell}/\CA_\ell\right]\right)^{-1} = \hat K^\text{IR}_\ell-\frac{\tilde \chi_\ell^2}{-C_\ell^{-1}-F +\tilde G_\ell}\ ,
\eqn{Kmatdef}
\eeq
where $\hat K^\text{IR}_\ell$ is given in \eq{KIR}, and we have defined the two other IR quantities
\beq
\tilde \chi_\ell\equiv \frac{1}{2\text{Im}[y]}\ ,\qquad \tilde G_\ell \equiv -\frac{M k^{2\ell+1}}{4\pi}\frac{{\rm Im}[z]}{{\rm Im}[y]}\ .
\eeq
Then the $C$-matrix may be written as
\beq
 {\mathbf C_\ell} = \left[ -F+ \tilde G_\ell+\frac{\tilde \chi_\ell^2}{\hat K_\ell - \hat K_\ell^{\rm IR}} \right]^{-1}\ ,
\eqn{Cmat2}
\eeq
This is our master equation for the $C$-matrix, the analog of \eq{Cmat1} for when both short- and long-range interactions are present in the effective theory. It depends on the regularization scheme through $F$, on the actual scattering data through $\hat K_\ell$, and on the $y$ and $z$ quantities numerically determined from the known IR physics through $\tilde \chi$, $\tilde G$ and $\hat K_\text{IR}$.   All quantities in the $C$-matrix    are manifestly real and are   even in $k$.  Furthermore, everything is finite: this is a fully renormalized theory given the initial assumption that the IR interaction is less singular than $1/r^2$.   When the long range interactions are turned off one has $\hat K_\ell^\text{IR}=0$, $\tilde\chi_\ell^2 = 1$ and $\tilde G_\ell= 0$ so that  the $C$-matrix of \eq{Cmat1} is recovered. 

We will generally consider two subtraction schemes, depending on which is more convenient:  the $\msbar$ scheme which sets $F=0$, and $\amsbar$ scheme which moves the lowest lying pole of the $C$-matrix to  $k=0$, meaning that $F$ is defined as
\beq
F = \lim_{k\to 0} \left[ \tilde G_\ell+\frac{\tilde \chi_\ell^2}{\hat K - \hat K_\ell^{\rm IR}} \right]\ , \qquad (\amsbar)
\eqn{Fdef2}\eeq
which is the generalization of \eq{Fdef1} to include the effects for long distance physics.   

We   give an example below of how this master equation works in an exactly solvable model before turning to the more complicated case of nucleons interacting via pions.  First, though, we briefly outline the formalism for coupled channels.

\subsection{Coupled channels}

The previous equations generalize to coupled channels with some minor adjustments (see, for example \cite{newton2013scattering}). Firstly, the Schr\"odinger equation is now a matrix equation. We will assume that parity is conserved, that $V$ is symmetric, and that the Wronskian is defined with a transpose and the following ordering
\begin{equation}
    W(F,G)=F^TG'-F'^TG.
\end{equation}
In what follows, $L$ is the diagonal matrix with the appropriate partial wave values along the diagonal. We borrow the notation used in \cite{newton2013scattering} that expressions such as $k^L$ are to be interpreted as the matrix with $k$ raised to the appropriate powers along the diagonal. The regular and irregular solutions are now defined as 
\begin{align}
    \mathcal{J}(r)&\equiv(F_L^+(k,r)e^{-i\pi L/2}y+c.c.)\\
    \mathcal{N}(r)&\equiv(F_L^+(k,r)e^{-i\pi L/2}z+c.c.)
\end{align}
where the Jost solutions are defined asymptotically to have the same plane wave factors along the diagonal with vanishing off diagonal components, and $y$ and $z$ are now matrices as well. The $S$ matrix for the purely long range potential is
\begin{equation}
    S^{\text{IR}}=-e^{-i\pi L/2}y(y^*)^{-1}e^{i\pi L/2},
\end{equation}
and is unitary and symmetric.
The normalization of our solutions may be found by promoting $\ell$ to $L$ and interpreting appropriately. The relation between the $y$ and $z$ matrices generalizes to
\begin{equation}
    y^{\dagger}z-y^Tz^*=-(i/2\mathbb){1}.
\end{equation}
The complete wavefunction solution is then given by $\Psi=\mathcal{J} A+\mathcal{N}B$, where $A$ and $B$ are real matrices. We choose to define \begin{equation}
    \hat{K}=\frac{4\pi i}{Mk}k^{-L}e^{i\pi L/2}(S-1)(S+1)^{-1}e^{-i\pi L/2}k^{-L}
\end{equation}
so that we may construct the $\hat{K}$ matrix as 

\begin{align}
\hat{K}&=\hat{K}^{\text{IR}}-\tilde\chi^T(-C^{-1}-F+\tilde{G})^{-1}\tilde\chi.
\end{align}
with
\begin{align}
    \hat{K}^{\text{IR}}&=\frac{4\pi}{Mk}k^{-L}(\mathrm{Re}y)(\mathrm{Im}y)^{-1}k^{-L},\\
    \tilde\chi&=k^{L}(2\mathrm{Im}y)^{-1}k^{-L},\\
    \tilde{G}&=-\frac{Mk}{4\pi}k^{L}(\mathrm{Im}y)^{-1}(\mathrm{Im}z) k^{L},
\end{align}
and $C$ identified with
\begin{equation}
    C^{-1}+F=\frac{Mk}{4\pi}k^{L}AB^{-1}k^{L}.
\end{equation}
Our $C$-matrix master equation for coupled channels is thus
\beq
    C=\left[
    -F+\tilde G +\tilde \chi \frac{1}{\hat K -{\hat K}^\text{IR}}
    {\tilde \chi}^T
    \right]^{-1}.
\eeq

 \subsection{A toy model with two $\delta$-shell potentials}
 
 To understand the appropriate power counting for dimer effective field theory in the presence of long range interactions, we begin with a simple but instructive toy model, consisting of one UV $\delta$-shell and one IR $\delta$-shell,
 \beq
   V(r)=V_\text{UV}(r) + V_\text{IR}(r)\ ,\qquad  V_\text{UV}(r)=-g\frac{\Lambda_\chi}{M}\delta(r-1/\Lambda_\chi)\ ,\qquad V_\text{IR}(r)=- \left(\frac{m_\pi}{\Lambda_{NN}}\right)\frac{m_\pi}{M}\delta(r-1/m_\pi)\ .
 \eeq
 $V_\text{UV}$ is an attractive  $\delta$-shell potential at the UV radius $1/\Lambda_\chi$ with strength $g$, where $\Lambda_{\chi}$ plays the role of the cutoff of the chiral Lagrangian;  $V_\text{IR}$ is an attractive  $\delta$-shell potential at radius $1/m_\pi$ with strength $m_\pi/\Lambda_{NN}$, like the real one pion exchange potential.  In the $S$-wave, either of the $\delta$-shell potentials by themselves would exhibit a bound state if their coupling were greater than one ($g$ for $V_\text{UV}$, $m_\pi/\Lambda_{NN}$ for $V_\text{IR}$.)

We have already seen that in the absence of $V_\text{IR}$, the $C$-matrix describing $V_\text{UV}$ would be given by \eq{CmatUV}.   When long-range interactions are included, we would expect that the  $C$-matrix should remain unchanged, since in the effective theory the contact interactions are supposed to encode the effects of $V_\text{UV}$, while the effects of $V_\text{IR}$ are included explicitly.  While this claim  is not immediately obvious from the formula in \eq{Cmat2}, it is in fact true and follows immediately from the relation which can be proved using results from the previous section that the modified $K$-matrix for this system is given by
\beq
    \hat{K}_\ell=\hat{K}_\ell^{\text{IR}}-\frac{\tilde{\chi}_\ell^2}{-(\hat{K}_\ell^{UV})^{-1}+\tilde{G}_\ell} \, ,
\eqn{Krel}\eeq
where $K_\text{UV}$ is given in \eq{KUV}, $\hat{K}_\ell^{\rm IR}$ is given by the same expression with the substitutions $g\to m_\pi/\Lambda_{NN}$ and $\Lambda\to m_\pi$, $\xi\to k/m_\pi$:
\beq
 \hat{K}^{\rm IR}_\ell = -\frac{4\pi}{M \Lambda_{NN} k^{2\ell}}\frac{ j_\ell(k/m_\pi)^2}{1+  (k/\Lambda_{NN})j_\ell(k/m_\pi)y_\ell(k/m_\pi)}\ .
\eeq
The other IR quantities may be computed straightforwardly with the result
\beq
     \tilde{\chi}_\ell=\frac{1}{1+  (k/\Lambda_{NN})j_\ell(k/m_\pi)y_\ell(k/m_\pi)}\ ,\qquad
  \tilde{G}_\ell=-\frac{Mk^{2\ell+1}}{4\pi}\left(\frac{k}{\Lambda_{NN}}\right)\frac{y_\ell^2(k/m_\pi)}{1+  (k/\Lambda_{NN})j_\ell(k/m_\pi)y_\ell(k/m_\pi)}\ .
 \eqn{Chishell}\eeq
 On the substitution of \eq{Krel} into \eq{Cmat2} one recovers \eq{Cmat1}, which shows that the $C$-matrix is determined entirely by the UV physics:  

\beq
{\mathbf C_\ell}  = \left[ \hat K_\text{UV}^{-1} - F\right]^{-1}\ ,\qquad F = \lim_{p\to 0} \hat K_\text{UV}^{-1} \ .
\eeq
Thus the $C$-matrix for $V = \left(V_\text{UV} + V_\text{IR}\right)$ is exactly the same as that which we obtained in \eq{CmatUV} for $V = V_\text{UV}$.     This result follows cleanly in the present example because $V_\text{UV}$ and $V_\text{IR}$ do not overlap in space.  When we look at the realistic one-pion exchange potential in the spin-triplet channels, this separation becomes renormalization scale  dependent.

\section{The procedure for determining dimers from data}
\label{sec:Fitting}

The real problems of interest are of course more difficult to deal with than the examples we have considered. We are not starting from a solvable model with known UV physics and an amplitude we can compute and analytically continue into the complex plane; a key part of the $C$-matrix is $\hat K$ which depends on data collected on the real momentum axis and no obvious way to treat it as a function that can be analytically continued to complex $k$; we are dealing with potentials more singular then $r^{-2}$ at the origin, which need to be regulated and renormalized.  Above all, we don't want to blindly try to fit data on the real axis to a theory with any number of dimers without having a good idea how many dimers are required, roughly where their pole positions are in the complex plane, and roughly what their couplings are (the residues at the poles).  We need to devise a method for peering into the complex momentum plane to get an idea of what is required by the physics. Before turning to the actual pion potential and real data, we sketch out the procedure we use to overcome the obstacles and achieve this objective. Key to our approach is the prejudice that the poles in the $C$-matrix obstructing the Weinberg power counting arguments do not arise solely from UV physics.

\begin{figure}[t]
    \centering
    \includegraphics[width=0.33\linewidth]{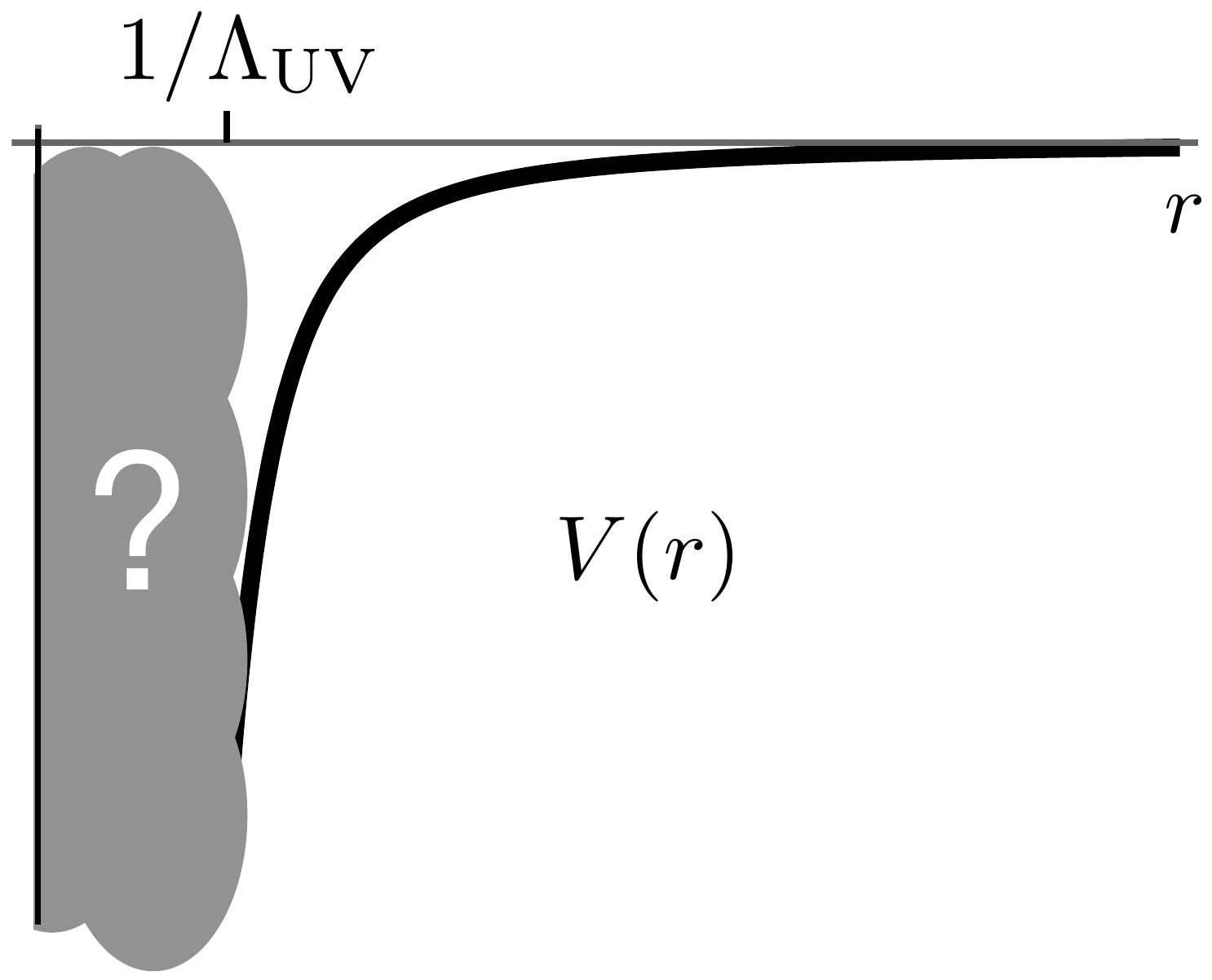}
    \caption{\it A cartoon of the problem we are interested in:  a potential with known behavior at large-$r$ but unknown at short distance. }
    \label{fig:Vreal}
\end{figure}

Suppose we have a well defined potential $V(r)$ (e.g. the one pion exchange potential) that gets the long distance behavior of the true interaction mostly right, but we don't trust it at all at distances shorter than $1/\Lambda_\text{UV}$, as sketched in Fig.~\ref{fig:Vreal}.  The effective theory we wish to construct consists of particle exchange accounting for $V(r)$ plus a good approximation of the $C$-matrix from  local operators, with whatever dimers are required at $O(Q^{-2})$ in the region $|k|\lesssim \Lambda_\text{UV}$ in the complex momentum plane, plus the usual expansion in contact terms up to the appropriate order one is working in the $Q$ expansion, which is now expected to involve powers of $Q/\Lambda_\text{UV}$ or smaller.  Unlike the contact interactions which can be fit to data on the real axis, the dimers require knowledge about the $C$-matrix in the complex plane where we have no data.  To solve this problem, the first step is to create a model potential that has the same  long-range  behavior below $\Lambda_\text{UV}$ as $V(r)$ but with  UV behavior regulated in a particular way.   We will use this crude model to (i) investigate its $C$-matrix to decide whether we expect the true potential to exhibit low-lying poles, and if so (ii) to determine their locations and residues for the model potential to be used as starting points for a fit to the physical data.   To that end we define the model potential $U_+(r,\mu_+)$ which is a regulated version of the original potential, defined as
\beq
U_+(r,\mu_+) = V(r)\theta(r-1/\mu_+)
\eeq
where we adjust $\mu_+$ to get the predicted scattering data as similar to the actual data as possible in the $k\lesssim \Lambda_{NN}$ region; $\mu_+$ should be roughly comparable to $\Lambda_{UV}$, and well above $\Lambda_{NN}$.  This will not be a great fit, and it is nothing close to the final prediction for the dimer effective theory; it is simply a starting point for peeking at the behavior in complex plane  of an approximate $C$-matrix for a potential that gets the long-distance parts of the actual potential correct.  Our prejudice is that even if this model has the wrong UV physics and doesn't fit the data well, it does get the region of the angular momentum barrier hump at $r\approx \Lambda_{NN}^{-1}$ right, where we believe lies much of the crucial physics  Weinberg's expansion doesn't account for correctly.

\begin{figure}[t]
    \centering
    \includegraphics[width=0.3\linewidth]{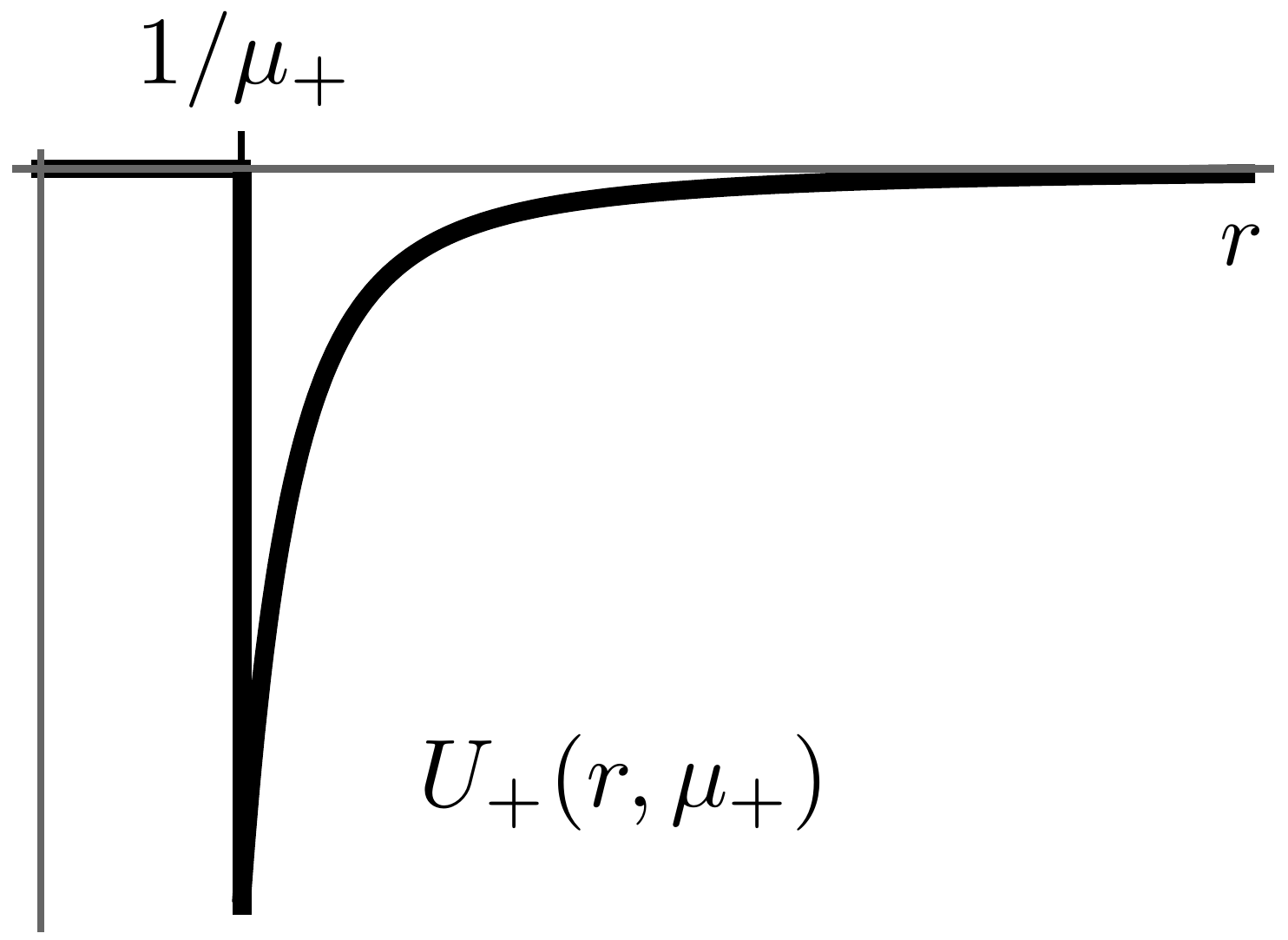}
 \hfill
   \includegraphics[width=0.3\linewidth]{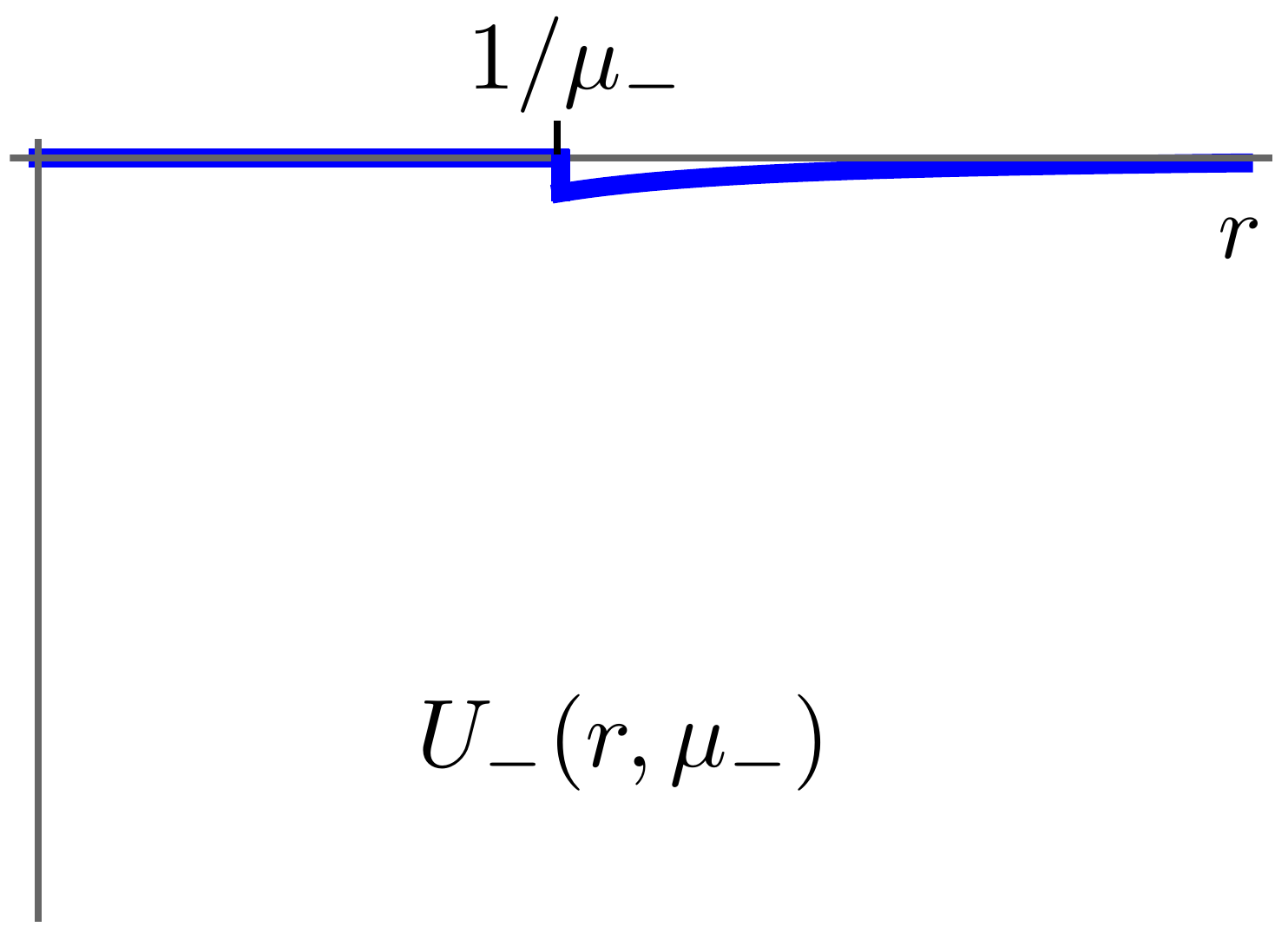}
 \hfill
   \includegraphics[width=0.3\linewidth]{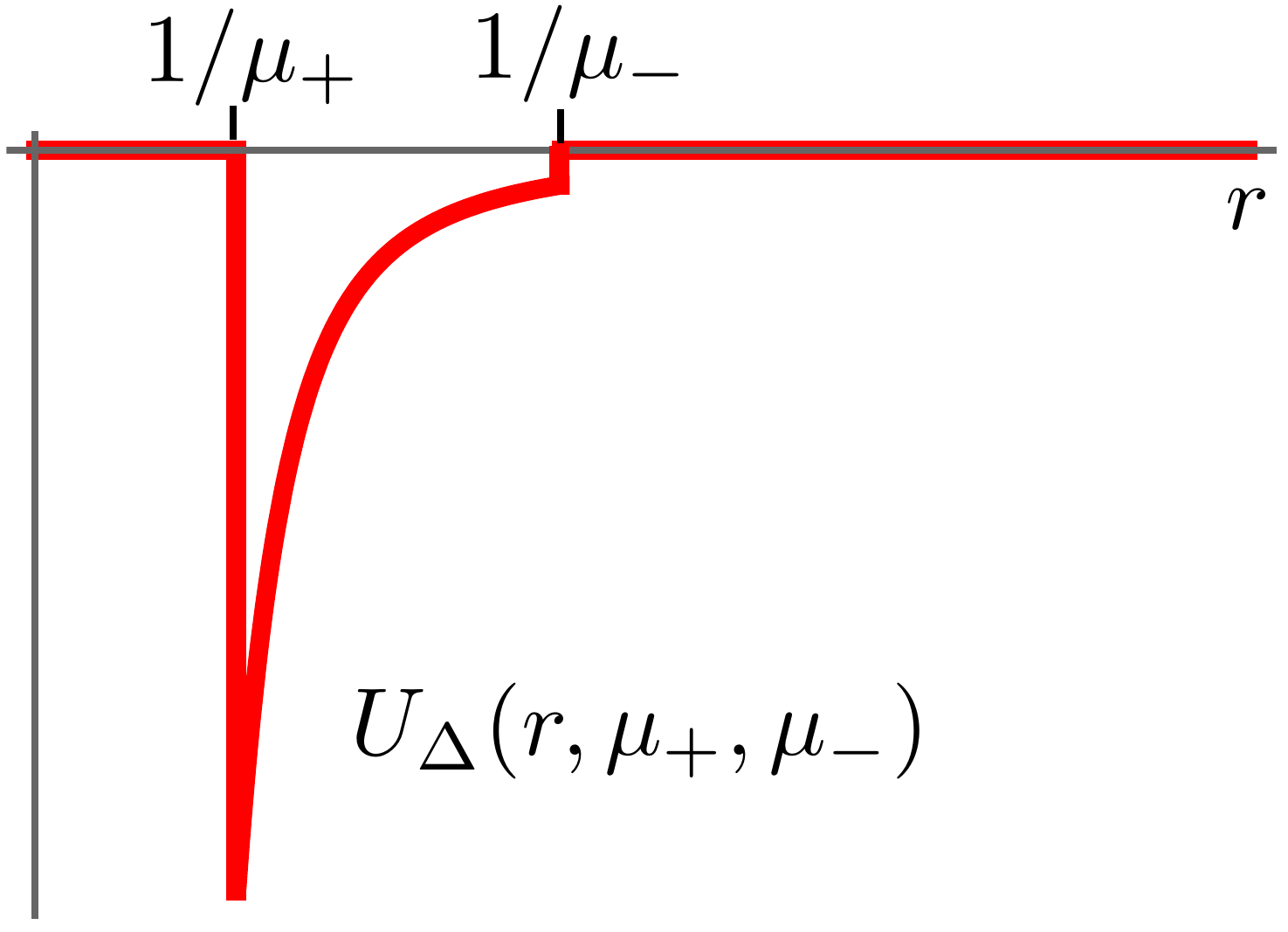}

   \caption{\it The potentials $U_+$, $U_-$ which are regulated versions of $V(r)$ with high and low cutoffs respectively, while $U_\Delta$ is their difference. With $V(r)$ given by the one pion exchange potential, the scale $\mu_+$ is chosen to optimize the fit to the actual data. }
    \label{fig:Upot}
\end{figure}

We also consider the potential
\beq
U_-(r,\mu_-) = V(r)\theta(r-1/\mu_-)
\eeq
which again is a regulated version of the potential of interest $V(r)$, but now  with a very low UV cutoff $\mu_-$, at the pion mass, for example, chosen so that $\mu_- < \Lambda_{NN}<\mu_+$.  The difference between these two potentials is called $U_\Delta$: 
\beq
U_\Delta(r,\mu_+,\mu_-) =U_+(r,\mu_+) -U_-(r,\mu_-) = V(r)\Bigl[\theta(r-1/\mu_+)-\theta(r-1/\mu_-)\Bigr]\ .
\eeq
These three potentials $U_+,U_-,U_\Delta$ are pictured in Fig.~\ref{fig:Upot}.   We now use the fact that $U_+$ is exactly given by $U_- + U_\Delta$ and that we can construct the effective theory for this model treating $U_-(r,\mu_-)$ exactly as our ``$V_\text{IR}(r)$'' and $U_\Delta(r,\mu_+,\mu_-) $ as our $V_\text{UV}(r)$, which is to be replaced by contact interactions by expanding the relevant $C$-matrix following the procedures in \S\ref{sec:KCIR}.  The formula in \eqs{Cmat2}{Fdef2}  requires computing $\tilde \chi$, $\tilde G$ and $\hat K_\text{IR}$ from the Schr\"odinger equation involving $V_\text{IR} = U_-$, as well as $\hat K$ from $(V_\text{IR} + V_\text{UV})=U_+$...but we do not need to do all that work.  The lesson learned in \S\ref{sec:KCIR} was that all these quantities conspire to make the $C$-matrix only involve $V_\text{UV}$, which is $U_\Delta$ in our case.  Therefore all we have to compute is the $\hat K_\text{UV}$ function associated with $U_\Delta$.
The reason why we can do this and represent the $C$-matrix in terms of dimers and contact interactions is that  $U_\Delta(r,\mu_+,\mu_-) $ is an exactly known potential which satisfies the prerequisite of \S\ref{sec:KCIR} that it vanish more rapidly than $r^{-2}$ at the origin (since it vanishes exactly for $r<\mu_+^{-1}$).  $U_\Delta$  also vanishes in the IR for $r>\mu_-^{-1}$, a property that guarantees that the scattering amplitude associated with $U_\Delta$ and hence the $C$-matrix will be meromorphic -- in particular, there will be no cuts in the complex plane \cite{newton2013scattering}.  Therefore one can systematically include dimers and contact interactions to reproduce it to any desired degree of accuracy out to any radius in the complex momentum plane... although in practice we will only dimerize the poles in the $C$-matrix out to  a radius $|k|\lesssim \mu_+$ and only compute contact interactions to  order $O(Q^0)$, the order of the one pion exchange potential. 

Note that if we succeed in computing this $C$-matrix out to the $\mu_+$ scale, this effective theory will represent our model -- where the nucleon-nucleon potential equals $U_+$ -- and will be accurate out to $|k|\lesssim \mu_+$, despite the fact that much of pion physics has been thrown away on replacing $U_\Delta$ with dimers and contact interactions\footnote{This claim may seem surprising but that is the lesson learned already from the initial example of a $\delta$-shell in \S\ref{sec:dshell1}.  One might think that cuts would be an obstacle, such as the $\ln(1+4k^2/m_\pi^2)$ term one finds at one-loop from one pion exchange, but the proof that the scattering amplitude from $U_\Delta$ will be meromorphic means that the relevant physics from any such cut in the nonperturbative treatment of the theory will be reproduced by the combination of $C$-matrix poles and the appropriate long distance quantities which correctly encode the Yukawa tail of the interaction.}.    To compute the desired $\hat K$ matrix we can numerically solve the Schrodinger equation with potential   $U_\Delta(r)$  for complex $k$ values, determine the two independent solutions $\CJ$ and $\CN$ for this complex energy,  extract the $y$ and $z$ coefficients defined in \eq{JNJost}, and compute the $\hat K$ matrix from \eq{Kmatdef}.  The only caveat is that the term ``${\rm Re}[\CA]$'' must be replaced by its analytic continuation off the real $k$ axis.

Once one has computed this $C$-matrix which replaces the potential $U_\Delta$ and has located its poles and residues, what does it mean and what does one do with it?  We work with the hypothesis that it provides a rough approximation of the $C$-matrix for the real problem, since it has captured correctly the physics we think is most important.  Therefore we include the same parameters in our effective theory for the real problem, letting their values float as we optimize the fit of this effective theory to the real-world data.  In most cases we find excellent fits with $\lesssim 30\%$ deviations from the values (pole positions and residues) we obtain from our toy model of Fig.~\ref{fig:Upot}. The appendix shows a more detailed comparison in two selected channels.

The dimer and contact interactions that we derive in this manner are associated with the renormalization scale $\mu_-$ but in this theory we can use exact renormalization group scaling to evolve the parameters back up to the $\Lambda_\text{UV}$ scale, as elaborated on in the following section. Every equation manipulated to obtain the $C$-matrix has used physical (renormalized) quantities, and so one would expect no dependence on renormalization scale.  However, as the $C$-matrix is evolved, poles move around in the complex plane, residues change, and one even sees that poles annihilate, merge, or are born.  If we had the full $C$-matrix to evolve, the predictions of the theory at different renormalization scales would be unchanged -- there would be no cutoff dependence.     However, when we truncate the expansion and only keep dimers out to a certain radius and only include a finite number of contact terms, then for example if a new pole quartet wanders into the $|k|\lesssim \Lambda_{UV}$ circle as $\mu$ is evolved, we have no parameter to take this physics into account.  As a result, the prediction for the phase shift exhibits some weak $\mu$-dependence and must be considered a source of uncertainty.  The error band in Fig.~\ref{fig:3P0phaseFinal.pdf} reflects such $\mu$ dependence.

\section{Application to $NN$ Scattering with pions}
\label{sec:NN}

We will now apply our program to NN scattering in the lowest angular momentum channels. Since we wish to describe scattering up to center of mass momenta of order 400 MeV, pion fields will be added explicitly into the theory, and we will treat pion exchange nonperturbatively. The leading order $O(Q^0)$ interaction comes from one pion exchange, giving us a potential matrix element
\begin{equation}
    V_{\text{OPE}}(\mathbf{q})=-\frac{g_A^2}{2f_\pi^2}(\mathbf{\tau_1}\cdot\mathbf{\tau_2})\frac{(\mathbf{q}\cdot\mathbf{\sigma}_1)(\mathbf{q}\cdot\mathbf{\sigma}_2)}{\mathbf{q}^2+m_\pi^2},
\end{equation}
which yields a coordinate space potential 
\begin{equation}
    V_{\text{OPE}}(\mathbf{r})=\frac{1}{3}(\tau_1\cdot\tau_2)\left[(\sigma_1\cdot\sigma_2)V_C(\mathbf{r})+S_{12}V_T(\mathbf{r})\right]
\end{equation}
where 
\beq
    V_C(\mathbf{r})=-\frac{4\pi}{M\Lambda_{NN}}\delta^3(\mathbf{r})+\frac{m_\pi^2}{M\Lambda_{NN}}\frac{e^{-m_\pi r}}{r}\ ,\qquad
    V_T(\mathbf{r})=\frac{m_\pi^2}{M\Lambda_{NN}}\left(1+\frac{3}{m_\pi r}+\frac{3}{(m_\pi r)^2}\right)\frac{e^{-m_\pi r}}{r}\ ,
\eeq
and $S_{12}=3(\sigma_1\cdot\mathbf{\hat{r}})(\sigma_2\cdot\mathbf{\hat{r}})-(\sigma_1\cdot\sigma_2)$.
The one pion exchange tensor potential acts only in spin triplet channels, and the one pion exchange contact interaction contributes only to $S$-wave scattering. The various isospin and spin matrix elements needed to perform our calculations are conveniently given for the lowest angular momentum channels in \cite{Birse:2005um}.

\subsection{Spin Triplet Channels}

We will begin with an analysis of the spin triplet channels, which are subject to the strong, singular interactions from the tensor part of one pion exchange. We exploit the same regulator introduced in the previous section, namely 
\begin{equation}
    V^{\text{reg}}_{\text{OPE}}(r)=V_{\text{OPE}}(r)\theta(r-1/\mu_-),
\eqn{Vreg}
\end{equation}
and perform calculations using this regulated potential. All scattering quantities then become dependent on $\mu_-$, and the $C$-matrix is constructed via
\begin{equation}
    C(\mu_-)=\left(\tilde{G}(\mu_-)+\frac{(\tilde{\chi}(\mu_-))^2}{\hat{K}-\hat{K}^{\pi}(\mu_-)}-F(\mu_-)\right)^{-1}.
    \eqn{Cmatmu}
\end{equation}
We again choose to work in the $\amsbar$ scheme, which shifts the primary dimer pole to the origin. This regulator and renormalization scheme is particularly convenient for calculations with the scattering formalism we have outlined, since the regular and irregular solutions may be computed unambiguously. The procedure of constructing the $\hat{K}$-matrix for the effective theory is straightforward. First, we choose a convenient $\mu_-$ and use the regulated potential in \eq{Vreg} to compute the asymptotic coefficients $y(\mu_-)$ and $z(\mu_-)$, and thus the long range functions $\hat{K}^{\pi}(\mu_-)$,  $\tilde\chi^2(\mu_-)$, and $\tilde{G}(\mu_-)$. From \eq{Cmatmu} and using real data for $\hat{K}$, we then compute the $C$-matrix in our chosen scheme on the real axis for this particular $\mu_-$. We expect this $C$-matrix to be well approximated by a meromorphic function with weak (higher order), unsubtracted cuts from two pion exchange, etc.,  in the data ($\hat K$). The previous section shows how we can use the pure one pion exchange potential for low values of $\mu_-$ to give a reasonable ansatz for the $C$-matrix, whose parameters we then fit to data. Since the primary dimer and subtraction constant can be fit or computed unambiguously, it is useful to subtract the primary dimer pole from \eq{Cmatmu} before further fitting. Once we have this fitted $C$-matrix, we can use it along with the long range functions to construct $\hat{K}$ for the effective theory using \eq{Kmatdef}, and compare to data. A powerful feature of this regulator is that we may relate the $C(\mu_-)$ we have just fit to the $C$-matrix at any other $\mu_+$ semi-analytically, using the pure one pion exchange potential. This is embodied in the exact RG equation \begin{equation}
    C(\mu_-)=\hat{K}^{\pi}_\Delta+\frac{C(\mu_+)\tilde{\chi}_\Delta^2}{1-C(\mu_+)\tilde{G}_\Delta}.\qquad\qquad (\msbar)
    \eqn{Cmatrel}
\end{equation}
A derivation of this equation is given in the appendix. As the notation is meant to suggest, the functions $\hat{K}^{\pi}_\Delta$, $\tilde{\chi}_\Delta$ and $\tilde{G}_\Delta$ are computed using the $U_\Delta$ regulated version of the one pion exchange potential of the previous section with UV and IR cutoffs given by $\mu_+$ and $\mu_-$ respectively (see $U_\Delta$ in Fig.~\ref{fig:Upot}). It is important to note that this equation is given in the  $\msbar$ ($F=0$) scheme; converting to the $\amsbar$ scheme is straightforward via the replacement $C^{-1}\to C^{-1}+F$. The utility of this equation is that we need only go through the process outlined in section IV and fit to real data once, at a value $\mu_-$ where we better understand what dimers are necessary in the theory. After using the RG equation to construct $C(\mu_+)$, we can recompute our long range functions at this new value of $\mu_+$ to construct $\hat{K}$ for the effective theory, regulated at $\mu_+$.  If this is done in small steps, we expect our $C$-matrix to change only slightly, with any cutoff dependence arising from our having thrown away poles beyond our range of interest (here defined as $|k|\lesssim 400$ MeV). The poles and residues of the $C-$ matrix at $\mu_+$ may be determined by computing all quantities in \eq{Cmatrel} in the complex plane, or by refitting (given that one knows the new dimer content). Both methods were utilized in our analysis, although the latter was used to generate more precise values. We perform this procedure within the range $\mu\in(100,1500)$ MeV in steps of $50$ MeV, keeping track of phase shifts showing the largest deviations from the data and whose curves are used to generate the cutoff bands. A more detailed procedure that may be followed to reproduce our results, as well as precise dimer parameters at convenient values of $\mu$ for all channels, are presented in the appendix.

\begin{figure}
    \centering
    \includegraphics[width=1.0\linewidth]{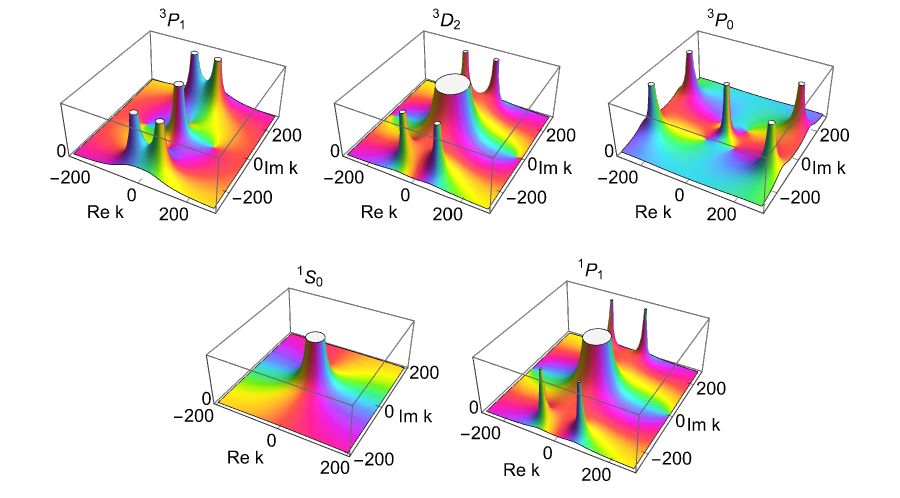}
    \caption{\textit{Top: $C$-matrices for the uncoupled spin triplet channels $^3P_1$, $^3D_2$ and $^3P_0$, computed in $\amsbar$ at chosen regulator momenta (see Table~\ref{tab:uncoupledparams} in the appendix) using the methods described in the text. Bottom: $C$-matrices for the spin singlet channels $^1S_0$ and $^1P_1$ in $\amsbar$. The former was computed without the cutoff regulator and the latter at $\mu=1$ GeV. As noted in the text, this $^1P_1$ $C$-matrix corresponds to the green curve in Fig.~\ref{fig:AllPhasesV2}. The $C$-matrix corresponding to the blue curve in Fig.~\ref{fig:AllPhasesV2} (a single primary dimer) is not pictured.}}
    \label{fig:Uncs}
\end{figure}

\begin{figure}
    \centering
    \includegraphics[width=1.0\linewidth]{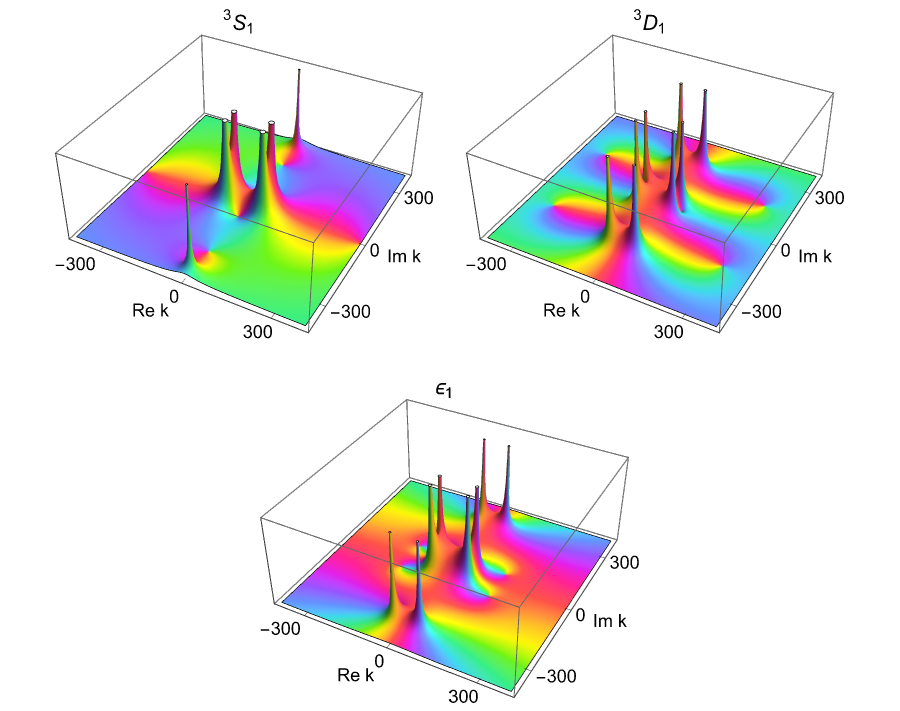}
    \caption{\textit{Here we show the $C$-matrix elements for the $^3S_1$$-^3D_1$ channel, computed at $\mu=140$ MeV in the $F=0$ scheme to show the similar pole structures. The resonance quartets are at identical pole locations (albeit with different residues) while the higher poles differ (discussed in the text).}}
    \label{fig:deuteronfriends}
\end{figure}

\begin{figure}
  \includegraphics[width=.90\linewidth]
  {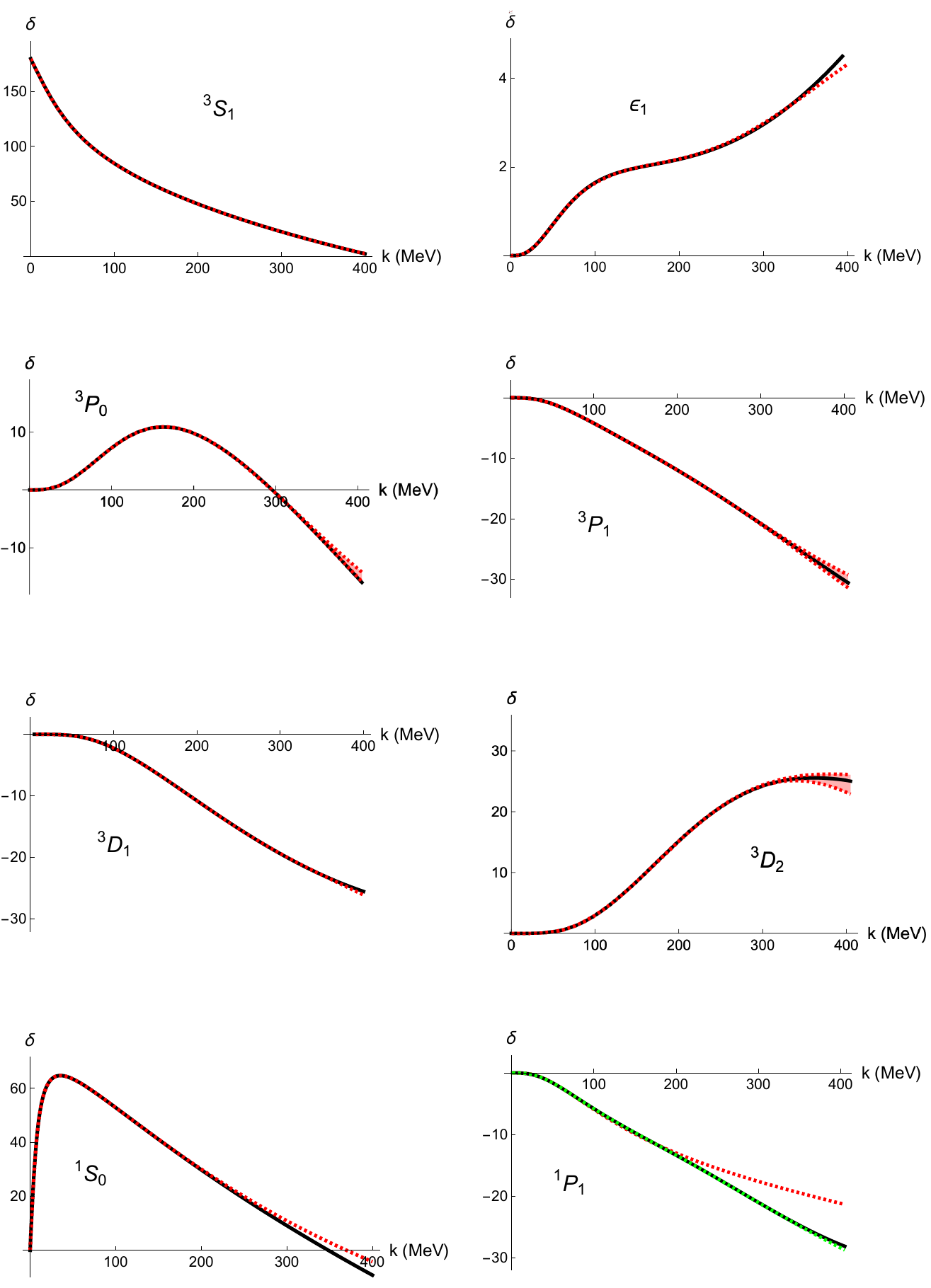}
   \caption{\textit{Phase shifts (degrees) versus momentum (MeV). The solid black curves are from the Nijmegen partial wave analysis while the dotted red curves show the $O(Q^0)$ calculations in the dimer effective field theory. These include one pion exchange and dimer content as pictured in Figs.~\ref{fig:Uncs},\ref{fig:deuteronfriends}. The bands in the uncoupled channels show the $\mu$ dependence of the calculation (from truncating the dimer expansion) in the range $\mu\in(100,1500)$ MeV. The dotted green curve in the $^1P_1$ channel shows the result of including a resonance quartet at $\approx200$ MeV in addition to the primary dimer; see \S~\ref{sec:p11} for a discussion.}}
     \label{fig:AllPhasesV2}
 \end{figure}

Now we present the results obtained from carrying out this procedure for $^3P_0$, $^3P_1$, $^3D_2$, and $^3S_1-$$^3D_1$. The $C$-matrices for the uncoupled channels (at chosen regulator momenta) are presented in the top panel of Fig.~\ref{fig:Uncs}. These $C-$matrices evidently contain additional resonance dimers parametrically at the scale $\Lambda_{NN}$. Using these $C-$matrices and the computed long range functions, we construct the phase shifts and compare to data from the Nijmegen partial wave analysis in Fig.~\ref{fig:AllPhasesV2}. The red bands in these figures show the effective $\mu$ dependence of the phase shift that results from truncating the dimer expansion. The band sizes are approximately $2^\circ$ and $4^\circ$ at 400 MeV in the $^3P_1$ and $^3D_2$ channels, respectively. In principle these bands would shrink if we allowed dimers beyond $\approx400$ MeV. The uncoupled channel fits show an improvement over the effective range expansion data and the usual leading order Weinberg calculation.  The fact that this is so seems to suggest that the additional non-analyticities from interactions like two pion exchange are quite weak (since we cannot account for these in the $C$-matrix purely from dimers), which is to be expected from power counting. Since the dimer is fit over a range of data, we are likely fitting subleading effects from interactions at the scale of $\Lambda_{NN}$. This would corroborate the 30\% change in values for the poles from pure one pion exchange. Naturally we would expect the peak where the singular potential begins to dominate to shift by powers of $\Lambda_{NN}/\Lambda_\chi$ as more singular potentials are added at higher orders in the expansion. These potentials will inevitably modify the locations of the dimers at higher orders, but this is naturally built into the dimer expansion via additional operators in the dimer Lagrangian. It would be interesting to carry out this procedure including the effects of two pion exchange, etc.  to verify these claims and push the radius of convergence of the expansion further. However, it seems promising that the dimers will be able to cure the issues observed with singular potentials, given that the IR physics is correctly accounted for and subtracted from the data. \par
In Figs.~\ref{fig:deuteronfriends}, \ref{fig:AllPhasesV2} we show the $C$-matrices and phase shift fits, respectively, for the coupled $^3S_1-$$^3D_1$ channel, which is also subject to an unnaturally large scattering length from the deuteron pole. We found it more convenient to work in the $F=0$ scheme for this channel. In addition to $C_0$ in the $S$ channel, we find strong evidence for a quartet of resonances at $\approx70$ MeV as well as a higher set of poles in each matrix element at $\approx300$ MeV for $\mu=140$ MeV where calculations were performed. Although the cutoff bands are not shown (they are difficult to compute for coupled channels), we observe similar cutoff dependence as in the uncoupled channels.

\subsection{Spin Singlet Channels} 
\label{sec:p11}

The spin singlet channels are markedly different than the spin triplets at leading order in the pion potential since, aside from the delta function in the $^1S_0$ channel which can be absorbed into the $C_0$ coefficient, they only feel the central, nonsingular part of one pion exchange. In particular, the lack of a singular power-law behavior of the potential near the origin means that the motivation for finding poles in the $C$-matrix inspired by Fig.~\ref{fig:hump} is not relevant for these channels, and indeed we find no evidence of resonance quartet poles.  Still, it is an elementary application of the dimer formalism for channels such as $^1S_0$, which has an unnaturally large scattering length due to the presence of a low lying virtual state in the amplitude, and a pair of poles in the $K$-matrix at low momenta. We then expect the primary dimer to play a significant role in this channel. Since the central potential has only a $1/r$ singularity, we can compute all long distance functions without the cuttoff regulator in $\msbar$, as in \cite{Kaplan:1996xu}. 


For the purposes of this calculation, we absorb the  contact interaction part of the one pion exchange potential  into the short distance $C$'s so that $C^{\text{eff}}_0=C_0+C_\pi$. After computing the long distance functions and fitting to data, we find no evidence for any additional $C$-matrix poles at higher momenta, so the $C$-matrix at $O(Q^0)$ is given by a single dimer at the origin and a contact interaction $C_0^{\text{eff}}$. This $C$-matrix is plotted in the lower panels of Fig.~\ref{fig:Uncs}, with the predictions for the phase shifts appearing in Fig.~\ref{fig:AllPhasesV2}, compared with Nijmegen data. The quality of the fit shown is comparable to the one achieved in the KSW scheme at NLO \cite{Kaplan:1998we}, although the iteration of one pion exchange included here does appear to improve agreement with the data. The inclusion of the dimer leads to a significant improvement over the LO and NLO Weinberg scheme fits (see \cite{Epelbaum:2008ga,Nogga:2005hy}).

In the same figures we also present the relevant quantities for the $^1P_1$ channel. At leading order in the $Q$ expansion the $C$-matrix only exhibits a central pole with no observed pole quartets and we only have only the primary dimer at the origin. This is consistent with the motivation discussed in the introduction, concerning the stationary state at the top of the angular momentum barrier since such a barrier does not exist in the spin-singlet channels using the one pion exchange potential.  Without the effects from singular one pion exchange, the power counting is as in the Weinberg scheme. Even with the inclusion of the dimer here, the LO calculation, although showing a reasonable fit, is not significantly better than the effective range expansion (or the calculation of pure one pion exchange without any additional $C$'s). This channel requires a significantly greater repulsion than is provided from one pion exchange alone. Whether or not this is fully remedied at higher orders in the chiral expansion isn't entirely clear, although it seems plausible that more singular potentials will be generated at higher orders in the chiral expansion, and these will again have the same pathologies as in the triplet channels (these effects, however, should be suppressed by powers of $Q/\Lambda_\chi$). If allow a fit to the $C$-matrix assuming a resonance at the expected scale in this channel, we obtain an improved agreement (even the undulations between 70 and 300 MeV are fit well, which is not true for the other fit) to the Nijmegen data. This seems to suggest that dimers will improve the agreement in singlet channels whose interactions become singular at higher orders, although the dimers will have additional suppression from powers of $\Lambda_\chi$ and will not enter until the appropriate order.  We stress that we do not take the quality of this fit as a success of our procedure, but simply as an indication that poles in the $C$-matrix may appear in the $\ell>0$ spin-singlet channels when two-pion exchange is included, which may again offer the dimers a way to improve agreement with data in these channels.

 \subsection{A coupled channel surprise}
 
All of the plots we give for the phase shifts assume that a local dimer theory exists to successfully account for the poles found in the $C$-matrix. The procedure discussed in \S\ref{sec:dimexp} shows how any quartet pole in an uncoupled scattering channel can be represented by a pair of dimers.  However, what does one do for coupled channels?
The obvious solution is to couple each dimer field to each of the fermion bilinear operators $(\psi\psi)_i$  with different $y_i$ couplings.  This immediately makes a couple predictions:  (i) in each $\{i,j\}$ element of the channels the pole location will be the same, and (ii) the residue in the $\{i,j\}$ coupled channel position for the pole in the first quadrant will be proportional to $y_i y_j$, which is a rank-1 matrix.  The $C$-matrices we find from data for the ${}^3S_1$-${}^3D_1$ coupled channels are pictured in Fig.~\ref{fig:deuteronfriends}.  We found these relations to be true for the innermost pole quartet:  they occur at the same positions in each channel and the residues satisfy 
$(Z_\text{SS} Z_\text{DD}) = (Z_{\epsilon_1})^2$ 
as predicted by the simple dimer construction.  However, this is patently not the case for the next farther out quartet, which are seen to not occur at the same position for the three partial waves\footnote{We note, however, that the existence of two sets of dimer poles at roughly the scales $70$ and $300$ MeV is consistent with the critical momentum values computed in \cite{Birse:2005um}. }.  Accounting for three different quartet pole locations will require three different sets of dimer fields.  If dimers behaved like ordinary particles and had hermitian interactions, this would not be possible:  to obtain a pole in the $\epsilon_1$ mixing channel one would need a dimer to couple to both the ${}^3D_1$ and the ${}^3S_1$ bilinears, and its pole would be seen then in the $SS$ and $DD$ components of the $C$-matrix as well. 

However, when all one requires is $\Phi$-hermiticity, it is possible to construct a theory where the ${}^3S_1$ bilinear can convert to $\Phi_\text{SD}$ which can subsequently convert to the ${}^3D_1$ without allowing the reverse processes to occur.  Thus $\Phi_\text{SD}$ exchange cannot contribute to the $SS$ or $DD$ components of the $C$-matrix.  Then a second field $\Phi_\text{DS}$ is introduced which does the reverse processes only, and $\Phi$-hermiticity exchanges the two fields.

This is not a particularly elegant solution -- in particular it does not explain why the three sets of quartet poles are rather close to each other.  However, it suffices to create a local effective theory that can account for the poles seen in the data and allow the derivative expansion to work up to much higher momentum.  A better understanding of the physics behind the behavior we see in Fig.~\ref{fig:deuteronfriends} would be interesting.

\section{Trimers and larger clusters}

A strong motivation for developing a robust and accurate effective theory for nuclear interactions is its ultimate application to nuclei and nuclear matter up to densities found in neutron stars.  It is possible that a successful effective field theory treatment of dense matter could require adding fundamental fields to the theory to represent clusters with baryon number three or higher. Although a detailed exploration of this possibility is beyond the scope of the present paper, we simply note here that the $C$-matrix formalism we have developed for the 2-body sector is readily generalized to higher body interactions.

\begin{figure}[h]
    \centering
    \includegraphics[width=0.7\linewidth]{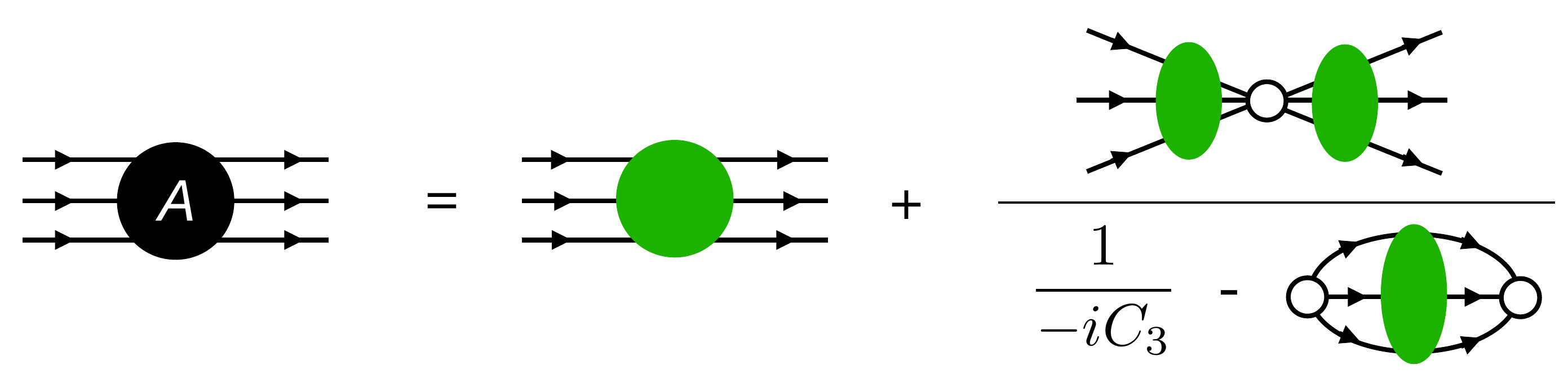}
    \caption{\it The 3-body scattering amplitude, consisting of the sum of 2-nucleon irreducible diagrams.  The blobs consist of all diagrams  involving interactions via pion exchange  or 2-body contact operators or $s$-channel dimer propagators, while $C_3$ represents all 3-body contact interactions and trimer exchange, in analogy to Fig.~\ref{fig:pibubbles}. }
    \label{fig:C3bubbles}
\end{figure}

In particular, the sum of diagrams shown in Fig.~\ref{fig:C3bubbles} gives rise to an expression analogous to the 2-body expression in \eq{master} for the 3-body scattering amplitude, which is qualitatively of the form (again, with the caveat that we ignore radiation pions)
\beq
\CA^{(3)} = \CA^{(3)}_\pi + \frac{\left(\chi^{(3)}\right)^2}{-\frac{1}{C_3} + G^{(3)}}\ .
\eeq
Here   $C_3$ is the 3-body contact interaction (including trimer exchange), $\CA^{(3)}_\pi$ is the scattering amplitude in the absence of 3-body contact interactions, $\chi^{(3)}$ can be interpreted as the 3-body wave function at the origin, and $G^{(3)}$ as the 3-body propagator from $\bfr = 0$ to $\bfr'=0$.  All of these quantities include the two-body contact terms as well as pion exchange.    As in the 2-body case, this can be inverted to give the $C_3$-matrix, where $\CA^{(3)}$ can be replaced by low-energy 3-body scattering data.  In general, the $C_3$-matrix will be meromorphic, and its poles can be represented by propagating fields representing trimer states with baryon number three.  This procedure can be readily generalized to clusters of any baryon number, so that $\alpha$ clusters in a nucleus such as ${}^{12}C$ could be represented by a tetramer, without any issue with an over-counting of degrees of freedom.  In dense matter, presumably many or most of these cluster states would develop large widths and dissolve, but their effects could still be significant as they would leave behind the $C$ coefficients for short range effects that were fit to few-body data where the clusters played a bigger role.

\section{Discussion}

We started this paper with a discussion for how effective theories for nuclear physics have had a  small radius of convergence for the momentum expansion, as compared to chiral perturbation theory for mesons,  and suggested that the explanation might be the existence of hidden nonanalytic structures in the complex $k$-plane associated with the scale $\Lambda_{NN}\approx 300\MeV$ which were obstructing the derivative expansion.  After pinpointing the $C$-matrix as the correct object to scrutinize for such obstructions in this nonperturbative theory, we indeed found poles at the expected scale in all of the spin-triplet partial waves for nucleon-nucleon scattering.  This motivated us to introduce propagating degrees of freedom to the theory -- so-called dimers -- that could account for this nonlocality explicitly, allowing the effective field theory expansion  in local operators to have a much larger radius of convergence, apparently up to the pion production threshold in most partial waves we examined\footnote{An interesting exception being the ${}^1P_1$ channel where we speculated that at higher order in the chiral expansion, the two-pion exchange potential might induce similar poles.}.  A feature of this approach is that the theory is fully renormalized and has no cutoff dependence. The way that the expected power law divergences manifest themselves is that as the renormalization scale $\mu$ is raised, there is an increasingly fine tuned cancellation between dimer parameters and pion exchange.  This makes it more difficult to analyze the theory at large $\mu$ but does not affect its predictions.  Small deviations  in predictions for different values of $\mu$ do enter the predictions from the interplay between the choice of renormalization scale and the truncation of the dimer expansion at some fixed radius from the origin of the complex $k$-plane.  Thus at one value of $\mu$ a pole in the $C$-matrix might be just outside the designated radius and not be represented by a dimer, while for a slightly different value of $\mu$ it will have entered the radius and a dimer will be introduced for it, leading to two slightly different predictions from two different values of $\mu$, despite the use of an exact renormalization group equation.

There are many open questions that have been barely touched on in this paper.  For example, how one can use this theory for studying the interactions of nucleons with photons and neutrinos or how it can be extended to the three-body sector or beyond, how to incorporate radiation pions,  whether mean field calculations can be performed in a theory with dimers, etc.

Perhaps the most important theoretical issues to address in order to make progress have to do with the regulator we used, which was to shut off the 2-body interaction at short interparticle separation.  This regulator -- which was  chosen for the ease of  implementation in the Schr\"odinger equation and for allowing us to make precise statements about the analytic properties of the amplitudes -- is difficult to incorporate at the Lagrangian level, which could be an obstacle for gauging the theory, implementing chiral symmetry, and so forth. Developing methods for constructing the $C$-matrix with regularization techniques that respect chiral symmetry and are more Lagrangian-friendly should be a priority, such as perhaps the gradient flow ideas  \cite{krebs2024toward,dbkgradflow1,dbkgradflow2} or Pauli-Villars regulators.

 The general formalism derived in this paper should be applicable to singular potentials in general, such as those relevant for atomic physics.
 
 \begin{acknowledgments}
 This research is supported in part by DOE Grant No. DE-FG02-00ER41132.  The authors also gratefully acknowledge the support of the University of Washington Royalty Research Fund.  We thank Silas Beane and Martin Savage for useful comments.
\end{acknowledgments}

\appendix

\section{The exact RG equation for $C(\mu)$}

In this appendix we provide a derivation of the renormalization group equation \eq{Cmatrel}, which provides an exact relation between the $C$-matrix at different renormalization scales.  Following the proof, we remark on how to use this equation in practice when one works with a truncation of the exact $C$-matrix.

We refer to the potentials $U_+$, $U_-$ and $U_\Delta$ in Fig.~\ref{fig:Upot}.  In the regions where these potentials vanish, the scattering solutions are linear combinations of the spherical Bessel functions $j_\ell$, $n_\ell$, while in the regions where the potentials are nonzero, the solutions are linear combinations of $\CJ_\ell$ and $\CN_\ell$ defined in \eq{JNnorm} from the Schr\"odinger equation with the IR potential $V(R)$.  At the boundary between the zero and nonzero regions  at radius $r=\mu^{-1}$ there is a $2\times 2$ transition matrix $T_\mu$ one can compute such that for a wave function 
\beq
\psi(r)  = \begin{cases} \alpha\, j + \beta \,n & r<\mu^{-1}\\ a \,\CJ + b \,\CN & r>\mu^{-1}\end{cases}\ , 
\eeq
the coefficients are related by
\beq
\begin{pmatrix} \alpha \\ \beta \end{pmatrix} = T_\mu \begin{pmatrix} a\\ b\end{pmatrix} \ .
\eeq
We now consider the IR potentials $U_+$ and $U_-$, along with UV physics represented by the $C$-matrices $C_+ = C(\mu_+)$ and $C_-=C(\mu_-)$ respectively, chosen so that the scattering waves $\psi_+$ and $\psi_-$ for the two cases correspond to identical $S$-matrices.  We seek the relation between $C_+$ and $C_-$, which will define our RG flow.   Constrained to give the same scattering matrix,  the two wave functions must be given by
\beq
\psi_\pm = \begin{cases} \alpha_\pm\, j + \beta_\pm\, n & r < \mu^{-1}_\pm \\ a\, \CJ + b\, \CN & r> \mu^{-1}_\pm\end{cases}\ ,
\eeq
sharing the same $a$ and $b$ coefficients so that they have the same asymptotic form.  From \eq{abc} we know that the  $C_\pm$ matrices are given in the $\msbar$ scheme ($F=0$) by
\beq
\frac{M k^{2\ell+1}}{4\pi}\,C_\pm \equiv \CC_\pm = \frac{\beta_\pm}{\alpha_\pm}\ .
\eeq
Therefore they will be related via 
\beq
\begin{pmatrix} \alpha_- \\ \beta_- \end{pmatrix} = T_{\mu_-} T^{-1}_{\mu_+} \begin{pmatrix} \alpha_+ \\ \beta_+ \end{pmatrix}\ ,\qquad\Longrightarrow \qquad
\CC_- = \frac{\left[T_{\mu_-} T^{-1}_{\mu_+}\right]_{21} +\left[T_{\mu_-} T^{-1}_{\mu_+}\right]_{22} \CC_+}{\left[T_{\mu_-} T^{-1}_{\mu_+}\right]_{11} +\left[T_{\mu_-} T^{-1}_{\mu_+}\right]_{12} \CC_+}\ .
\eqn{CCrel}\eeq
The combination of transition matrices $T_{\mu_-} T^{-1}_{\mu_+}$ may be conveniently related to the asymptotic properties of the scattering solutions  for the ``slice'' potential $U_\Delta$ in Fig.~\ref{fig:Upot}.  Such solutions look like
\beq
\CJ_\Delta = \begin{cases} j & r <\mu^{-1}_+ \\
c \CJ + d\CN & \mu^{-1}_+ < r < \mu^{-1}_- \\
\gamma j + \delta n & \mu^{-1}_- < r
\end{cases}\ ,\qquad
\CN_\Delta = \begin{cases} n & r <\mu^{-1}_+ \\
c' \CJ + d'\CN & \mu^{-1}_+ < r < \mu^{-1}_- \\
\gamma' j + \delta' n & \mu^{-1}_- < r
\end{cases}\ .
\eqn{JNDelta}
\eeq
 Their asymptotic properties are given by the $y_\Delta$ and $z_\Delta$ numbers relating the $\CJ$- and $\CN$-type solutions to the Jost solutions, as in \eq{JNJost}. We can read off from \eq{JNDelta} that these are given by
 \beq
 y_\Delta = -\half( i\gamma + \delta)\ ,\qquad z_\Delta = -\half( i\gamma' + \delta')\ .
 \eeq
Since these parameters can be defined via the transition matrices as
 \beq
 \begin{pmatrix} \gamma \\ \delta \end{pmatrix}  = T_{\mu_-} T^{-1}_{\mu_+}\begin{pmatrix} 1\\ 0 \end{pmatrix}\ ,\qquad 
\begin{pmatrix} \gamma' \\ \delta' \end{pmatrix}  = T_{\mu_-} T^{-1}_{\mu_+}\begin{pmatrix}0\\1 \end{pmatrix} \ ,
 \eeq
we can solve for the transition matrices in terms of $y_\Delta $ and $z_\Delta $:
\beq
T_{\mu_-} T^{-1}_{\mu_+}   = \begin{pmatrix}\gamma & \gamma' \\ \delta & \delta' \end{pmatrix} = -2 \begin{pmatrix}{\rm Im}[y_\Delta] & {\rm Im}[z_\Delta] \\{\rm Re}[y_\Delta] & {\rm Re}[z_\Delta]\end{pmatrix} \ .
%
%
%
\eeq
Inserting this solution into \eq{CCrel} we obtain
\beq
\CC_- = \frac{{\rm Re}[y_\Delta] +{\rm Re}[z_\Delta] \CC_+}{{\rm Im}[y_\Delta] +{\rm Im}[z_\Delta] \CC_+}\ .
\eqn{CCrel2}\eeq
Finally, using the definitions from \S\ref{sec:KCIR}
\beq
\hat K_\Delta = \frac{4\pi}{M k^{2\ell+1}} \frac{{\rm Re}[y_\Delta]}{{\rm Im}[y_\Delta]}\ ,\qquad
\tilde G_\Delta = - \frac{M k^{2\ell+1}}{4\pi} \frac{{\rm Im}[z_\Delta]}{{\rm Im}[y_\Delta]}\ ,\qquad
\tilde \chi_\Delta = \frac{1}{2{\rm Im}[y_\Delta] }\ ,
\eqn{KGCdef}
\eeq
along with the Wronskian relation $(y_\Delta^* z_\Delta- z_\Delta^* y_\Delta) = -i/2$ from \eq{Wyz} we obtain the desired result relating the $C$-matrices at the two different scales $\mu_+$ and $\mu_-$ in the $\msbar$ scheme:
\beq
C(\mu_-) = \hat K_\Delta + \frac{(\tilde \chi_\Delta)^2 C(\mu_+)}{1 - \tilde G_\Delta C(\mu_+)}\ .
\eeq
All quantities are to be computed in a particular partial wave, but we have dropped the $\ell$ subscripts for clarity.

The RG equation is exact, and if we had complete knowledge of the $C$-matrix at one renormalization scale $\mu_-$, we could use this equation to obtain the exact $C$-matrix at a higher renormalization scale $\mu_+$ by solving  the Schr\"odinger equation for the potential $U_\Delta(r)$ and from it determining the quantities $\hat K_\ell^\Delta$, $\tilde G^\Delta_\ell$, and $\tilde \chi^\Delta_\ell$ as defined by \eq{KGCdef} and the asymptotics of \eq{JNJost}.  A feature of this equation is that $U_\Delta(r)$ has both a UV cutoff ($\mu_+$) and an IR cutoff ($\mu_-$) and so all of these quantities are well defined and may be solved numerically in  the complex $k$ plane where the corresponding amplitude is necessarily meromorphic -- in particular, it lacks any cuts. 

\begin{figure}[t]
    \centering
    \includegraphics[width=0.35\linewidth]{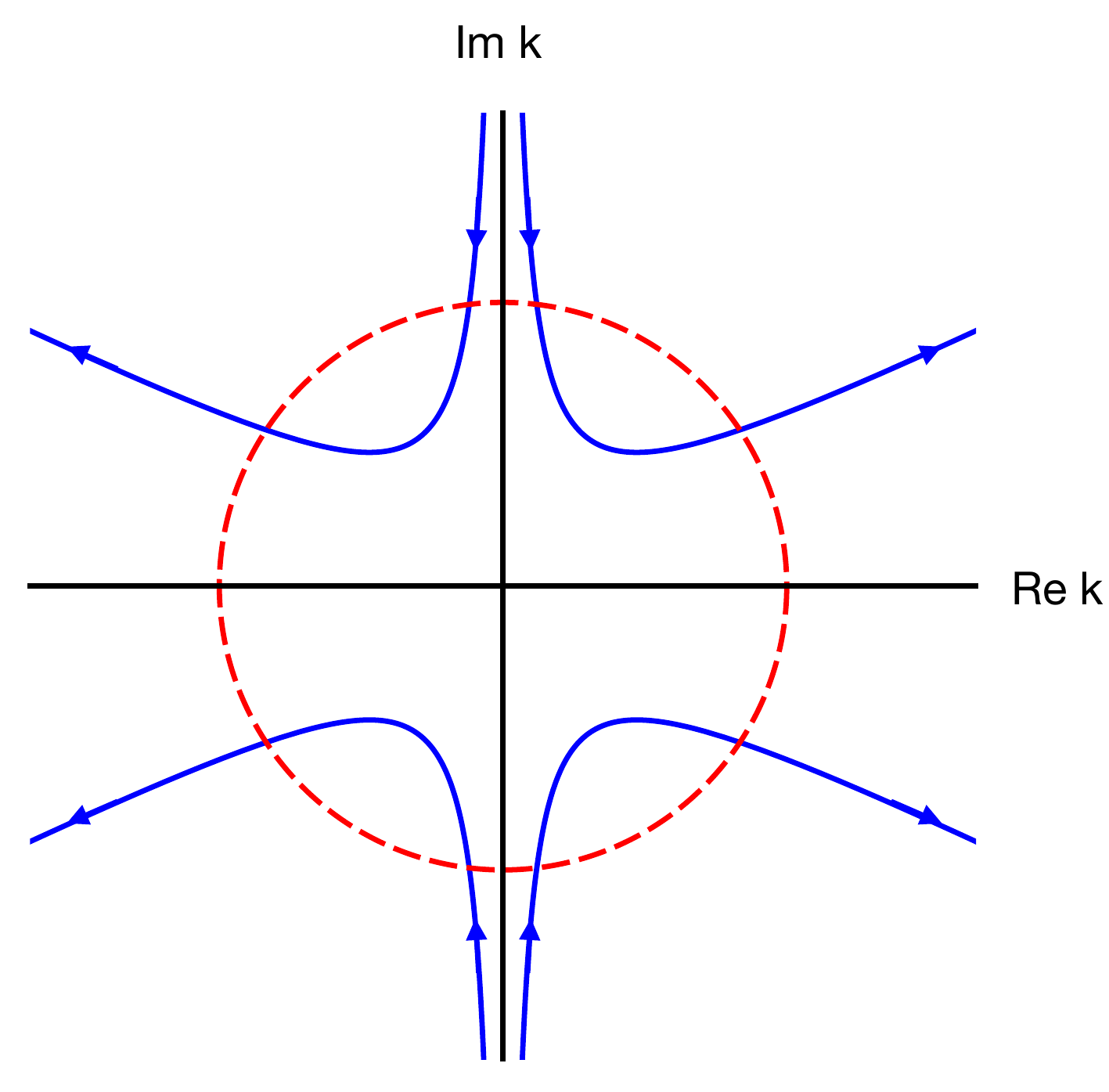}
    \caption{\textit{Hypothetical RG flow of a pole quartet in the complex momentum plane (blue lines) entering and leaving the region of convergence for the dimer effective theory  (dashed red circle).}}
    \label{fig:poleflow}
\end{figure}

In practice, however, one will be constructing the dimer effective field theory out to some desired radius of convergence (set to 400 MeV in this paper), representing poles in the $C$-matrix within that radius by dimers and ignoring the poles outside the specified radius.  This is not an RG-independent procedure, however. As we have emphasized, the $C$-matrix is not itself a physical quantity like the $S$-matrix, and the locations of its poles  flow with $\mu$.  In particular, as one changes $\mu$, poles in the $C$-matrix will flow in and out of the target radius of convergence, as depicted in Fig.~\ref{fig:poleflow}.  As the poles enter or leave the target radius, the correct thing to do is to add or remove appropriate dimers to the effective theory.  In this way, the Lagrangian one is working with will change discontinuously with $\mu$.  Depending how close the nearest poles are outside the target radius, which are not being accounted for with dimers, the resulting theory will give $\mu$-dependent answers.  The variation with $\mu$ will, however, always correspond in magnitude to effects one will expect to find from structures outside the designated convergence zone.  This is the $\mu$-dependence that is pictured in Fig.~\ref{fig:AllPhasesV2} for the uncoupled spin-triplet channels, and it provides a visual measure of typical errors being incurred in the truncation of the theory to a finite number of dimers.  We have not presented analogous results for the ${}^3S_1-{}^3D_1$ coupled channels, which would be more tedious to compute, but we see evidence that their behavior is similar to that of the uncoupled channels.

Similarly one risks incurring errors if one starts with a $C$-matrix only determined out to some radius at the initial value of $\mu$, and then use the RG equation to determine the $C$-matrix at a dramatically different value of $\mu$, since one may not be properly accounting for poles that have flowed into the target radius from far away.  The safest procedure is to track the evolution of the $C$-matrix through relatively small increments in $\mu$.

\section{Procedure and Parameters}
Now we outline the procedure which may be used to generate the phase shift plots and in addition the cutoff dependence incurred in calculations (the bands in the plots in section V).

\begin{figure}[t]
    \centering
    \includegraphics[width=0.70\linewidth]{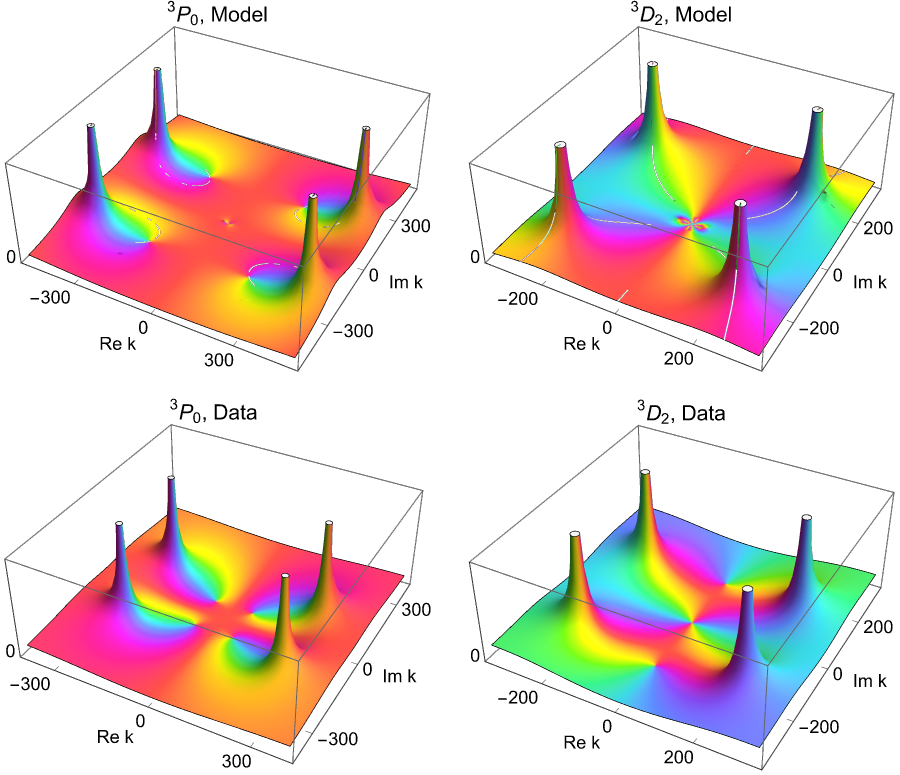}
    \caption{\textit{The top panel shows the functions $k^{2\ell}C(k^2)$ in the $^3P_0$ and $^3D_2$ channels. These were computed numerically with the one pion exchange potential whose UV cutoff was tuned to 686 MeV and 777 MeV in each of the respective channels to fit the data best. The IR cutoffs are taken at 120 MeV and 80 MeV, respectively. The bottom panel shows the same functions, now obtained by fitting to real data at the same $\mu$ scales. In the $^3P_0$ channel there is approximately a 30\% correction to the pole location and a 10\% correction to the residues, while in the $^3D_2$ channel the pole locations experience a 10\% correction and the residues a 20\% correction. All quantities are computed in $\amsbar$.}}
    \label{fig:OPEcomp}
\end{figure}

\begin{itemize}
\item Given a particular partial wave and IR potential, choose a convenient starting $\mu_-$ and construct the regulated potential via \eq{Vreg}.
\item Compute $y(\mu_-)$ and $z(\mu_-)$ using this potential, and thus $\hat{K}^{\pi}(\mu_-)$,  $\tilde\chi^2(\mu_-)$, and $\tilde{G}(\mu_-)$ across a grid of real momenta within the domain of interest (here $0-400$ MeV). 
\item Construct the $C(\mu_-)$ matrix via \eq{Cmatmu} with $F(p^2,\mu_-)$ defined to give the $C$-matrix a pole at the origin. Using either the modified effective range or by fitting, compute $C_{-2}$.
\item Subtract the dimer pole from this $C$-matrix so that it is smooth at the origin. This typically requires obtaining the residue to high precision. The next step is to determine if any additional dimers may be required to extend the radius of convergence. (Alternatively one may find it more convenient to work in the $F=0$ scheme for the fitting procedure, and then convert to the $\amsbar$ scheme when computing phase shifts). One may construct an ansatz for the residual $C$-matrix using the pure one pion exchange potential as a model, as described in section IV. This is illustrated in Fig.~\ref{fig:OPEcomp} for the $^3P_0$ and $^3D_2$ channels. By tuning $\mu_+$ to best fit the data and taking $\mu_-\approx m_\pi$ or lower, we can arrive at reasonable guesses for the pole locations and residues. By fitting to real data over the interval of interest using this guess, we can compute an approximate $C$-matrix. We observe typically no more than a 30\% correction if $\mu_\pm$ are chosen appropriately. The only caveat to this method is that $\mu_-$ has to be low enough to see reasonable agreement, but one in general observes more poles occurring in the model potential as $\mu_-$ is decreased. In any case, this method seems to give reasonable initial guesses.
\item Given an ansatz and initial guess for the residues and locations of the dimer poles, fit the $C$
matrix to \eq{Cmatmu} with the appropriate subtraction constant. The range of the fit should be high enough to capture information about the pole without sacrificing agreement at lower momenta. Typically fitting up to $300$ MeV satisfies each of these. \item Using the computed functions $\hat{K}^{\pi}(\mu_-)$, $\tilde\chi(\mu_-)$, $\tilde G(\mu_-)$, and now the fitted $C(\mu_-)$, construct the approximate $\hat{K}$ matrix and the phase shift from the defining relation \eq{Kmatdef}, and compare to data.
\item Use \eq{Cmatrel} to relate the $C$-matrix at a particular $\mu_-$ to another $\mu_+$. The appropriate way to use this equation is to extract from it the poles and residues of the new $C$-matrix (adding in any additional dimers that enter the range of interest). One may do this either by computing the functions $\hat{K}^{\pi}_\Delta$, etc, in the complex momentum plane (numerically costly but straightforward), or by computing the new $C(\mu_+)$ on the real axis and fitting it to a ratio of polynomials to identify pole locations. The latter method was used in conjunction with a fitting procedure once the new pole locations/types were identified.
\item Compute $\hat{K}^{\pi}(\mu_+)$, $\tilde\chi(\mu_+)$, $\tilde G(\mu_+)$ at the new value of $\mu_+$ and construct the $\hat{K}$ approximant at this new value, replotting the phase shift. 
\item Repeat the previous two steps in the desired range of cutoffs to compute the cutoff dependent bands (we use a range from 100-1500 MeV and steps of 50 MeV in our calculations. The curves with the largest deviations from this procedure are used to generate the figures). 
\end{itemize}

Dimer parameters at convenient initial values of $\mu$ are given in Tables~\ref{tab:uncoupledparams}, \ref{tab:coupledparams}. To reduce clutter we use the script $\mathcal{C}$-matrix to denote the $C-$matrix made dimensionless by multiplying by the appropriate factors of $\Lambda_{NN}$. In the uncoupled channels we have $\mathcal{C}=\frac{M}{4\pi}\Lambda_{NN}^{2\ell+1}C$ (and $\mathcal{F}_0=\frac{4\pi}{M}\Lambda_{NN}^{-2\ell-1}F_0$), while in the coupled channels the notation means the relevant matrix element after projecting by the matrices $\Lambda_{NN}^L$ on each side of the $C$-matrix, $\mathbf{\mathcal{C}}=\frac{M\Lambda_{NN}}{4\pi}\Lambda_{NN}^L\mathbf{C}\Lambda_{NN}^{L}$. The residues are given in a form so that the $\mathcal{C}$-matrix is constructed as $\mathcal{C}=\mathcal{C}_{-2}/p^2+\mathcal{C}_{r}/(p^2-p_*^2)+\cdots$, so that both $\mathcal{C}_{-2}$ and $\mathcal{C}_r$ have dimension $(\text{mass})^2$. We choose values of the regulator momenta where only a single set of dimers was required to achieve the fit. These may be used to construct the initial $C$-matrix (at that particular $\mu$), with the understanding that all possible reflections of poles be included in the sum with appropriate residues. In conjunction with the previous steps one may reconstruct the phase shifts as a function of $\mu$ in any range of interest. 

An important question is whether there is a unique dimer configuration to be used as the starting point in calculations. For example, for resonances close to the imaginary axis it may also be possible to achieve reasonable fits with a single dimer on the imaginary axis. Even with this change, the RG equation would have to reproduce the appropriate 
poles in $C$ for higher values of $\mu$ (especially when these occur on the real axis). Thus it may be plausible that different initial conditions for $C$ lead to ultimately the same/similar behaviors as $\mu$ is increased. 
\begin{table}[h!]
\centering
\setlength{\tabcolsep}{12pt}
\renewcommand{\arraystretch}{2.1} 
\begin{tabular}{c|c c c c c}
\hline
 & $^1S_0$ & $^3P_0$ & $^3P_1$ & $^3D_2$ & $^1P_1$ \\
\hline
$\mu\,\, (\text{MeV})$  & $140^*$ & $120$ & $220$ & $180$  & $1000$\\
$\displaystyle \mathcal{F}_0$ & -0.4124 & 5.345 & -1.488 & -1.848 &  3.619 \\
$\displaystyle \mathcal{C}_{-2}/\Lambda_{NN}^2$& -0.7103 & 5.155$\times10^{-3}$ & -2.398$\times10^{-2}$ & -5.373$\times10^{-2}$ & 6.424$\times10^{-2}$  \\
$p_*\,\, (\text{MeV})$ & - & -248.7-137.0$i$ & -55.2+234.0$i$ & -70.5-239.8$i$ &  -151.4-181.5$i$ \\
 $\displaystyle \mathcal{C}_r/\Lambda_{NN}^2$& - & 0.2701-0.1765$i$ & 0.1983-0.4607$i$ & 0.03678-0.04161$i$ & 0.1437-0.01537$i$ \\
 $\displaystyle \mathcal{C}_0$& 0.2930 & - & - & - & -  \\[1.2ex]
\hline
\end{tabular}
\caption{\textit{Dimer parameters that may be used to construct $C(\mu)$ for the uncoupled channels. All but the pole locations are given in dimensionless form, as discussed in the text. The asterisk on the 140 in the $^1S_0$ column indicates that it was computed in $\msbar$ as opposed to using the cutoff regulator. We use the values $M=939$ MeV, $m_\pi=140$ MeV, and $\Lambda_{NN}=290$ MeV for calculations.}}
\label{tab:uncoupledparams}
\end{table}
\vfill\eject
\begin{table}[t]
\centering
\setlength{\tabcolsep}{12pt}
\renewcommand{\arraystretch}{2.1} 
\begin{tabular}{c|c c c}
\hline
 & $SS$ & $SD=DS$ & $DD$ \\
\hline

$p_A$ (MeV)& $65.9+30.4i$ & $65.9+30.4i$ & $65.9+30.4i$ \\
$p_B$ (MeV)& $362.6i$ & $45.9+297.4i$ & $45.0+234.0i$ \\
$\displaystyle \mathcal{C}_{-2}^{(A)}/\Lambda_{NN}^2$& $-0.2642+0.5527i$ & $0.1473+0.5069i$ & $0.3818+0.2395i$  \\
$\displaystyle \mathcal{C}_{-2}^{(B)}/\Lambda_{NN}^2$& -0.8087 & $-0.1425+1.3466i$ & $-0.2987+1.7174i$ \\
$\displaystyle \mathcal{C}_0$& 0.5093 & - & - \\
 [1.2ex]
\hline
\end{tabular}
\caption{\textit{Dimer parameters for the $^3S_1$$-^3D_1$ channel at $\mu=140$ MeV in the $F=0$ scheme. The columns indicate the relevant matrix element (after projecting by the appropriate factors of $\Lambda_{NN}$ on each side), and the scripts $A$ and $B$ signify the innermost and outermost set of dimers, respectively.  }}
\label{tab:coupledparams}
\end{table}

\bibliography{dimer_refs.bib}
\end{document}